\documentclass[11 pt,nofootinbib,notitlepage,eqsecnum]{revtex4-1}

\usepackage[pdfencoding=auto, psdextra]{hyperref}
\usepackage[utf8]{inputenc}
\usepackage{amsmath,amssymb,amsthm}

\usepackage{scalerel,stackengine}
\stackMath
\newcommand\reallywidehat[1]{%
\savestack{\tmpbox}{\stretchto{%
  \scaleto{%
    \scalerel*[\widthof{\ensuremath{#1}}]{\kern-.6pt\bigwedge\kern-.6pt}%
    {\rule[-\textheight/2]{1ex}{\textheight}}
  }{\textheight}%
}{0.5ex}}%
\stackon[1pt]{#1}{\tmpbox}%
}

\newcommand{\pvec}[1]{\vec{#1}\mkern2mu\vphantom{#1}}

\DeclareMathOperator{\sgn}{\mathrm{sgn}\,}

\def\Re{\mathrm{Re}\,}
\def\Im{\mathrm{Im}\,}

\def\R{\mathbb{R}}

\def\p{\vec{p}}
\def\F{\vec{F}}
\def\f{\vec{f}}
\def\fom{\mathring{\omega}}
\def\fe{\mathring{e}}
\def\fq{\mathring{q}}
\def\tf{\tilde{f}}

\def\c{\vec{c}}
\def\b{\beta}
\def\l{\vec{\lambda}}
\def\sgnp{\overrightarrow{\sgn p}}

\newcommand{\dd}{\mathrm{d}}

\newcommand{\SU}{\mathrm{SU(2)}}

\newcommand{\hc}{\text{h.c.}} 
\newcommand{\cc}{\text{c.c.}} 

\newcommand{\BI}{\gamma}
\newcommand{\BO}{\mathcal{O}} 
\newcommand{\lp}{\ell_p} 

\def\beq{\begin{equation}}
\def\eeq{\end{equation}}
\def\bq{\begin{equation*}}
\def\eq{\end{equation*}}

\begin{document}

\title{Deriving loop quantum cosmology dynamics from diffeomorphism invariance}
\author{Jonathan Engle\footnote{jonathan.engle@fau.edu} and Ilya Vilensky\footnote{ilya.vilensky@fau.edu}}
\affiliation{Florida Atlantic University, 777 Glades Road, Boca Raton, FL 33431, USA}
\begin{abstract}
\centerline{\bf Abstract}
We use the requirement of diffeomorphism invariance in the Bianchi I context to derive the form of the quantum Hamiltonian constraint. After imposing the correct classical behavior and making a certain minimality assumption, together with a certain restriction to ``planar loops'',  we then obtain a unique expression for the quantum Hamiltonian operator for Bianchi I to both leading and subleading orders in $\hbar$. Specifically, this expression is found to exactly match the form proposed by Ashtekar and Wilson-Ewing in the loop quantum cosmology (LQC) literature. Furthermore, by using the projection map from the quantum states of the Bianchi I model to the states of the isotropic model, we constrain the dynamics also in the homogeneous isotropic case, and obtain, again to both leading and subleading order in $\hbar$, a quantum constraint which exactly matches the standard `improved dynamics' of Ashtekar, Pawlowski and Singh. 
This result in the isotropic case does not require a restriction to planar loops, but only the minimality assumption.
Our results strengthen confidence in LQC dynamics and its observational predictions as consequences of more basic fundamental principles. Of the assumptions made in the isotropic case, the only one not rigidly determined by physical principle is the minimality principle; our work also shows the exact freedom allowed when this assumption is relaxed.
\end{abstract}
\maketitle

\section{Introduction}

Loop quantum gravity (LQG) \cite{al2004, rovelli2004, thiemann2007, ap2017} is an approach to quantizing general relativity that is based on viewing gravity as a gauge theory with diffeomorphisms playing the role of gauge transformations. The framework of loop quantum cosmology (LQC) \cite{bojowald2008, as2011a, as2017} has been developed in order to provide observational predictions for the early universe and to test LQG-derived quantization techniques in the simplified symmetry-reduced cosmological setting. In particular, for homogeneous spacetimes almost all diffeomorphism symmetry is fixed except for a three-parameter family of \textit{residual diffeomorphisms}.

It is important to understand the choices and ambiguities in the quantization procedure both to establish the robustness of the physical theory and its predictions as well as to explore alternative quantizations in case one of them provides a better model of nature. In loop quantum gravity one follows the Dirac quantization procedure by defining a quantum algebra of basic observables (known as the holonomy-flux algebra) and finding a Hilbert space representation of this algebra. This part of the procedure is termed \textit{kinematical}. Then one imposes the constraints of general relativity as quantum operator equations and finds the solutions of the constraints in the kinematical Hilbert space or its dual. In particular, the Hamiltonian constraint generates time reparametrization invariance and therefore defines the dynamics of the kinematical states.

In the full theory a natural question is how the representation of the quantum algebra is chosen. In \cite{lost2006} the authors proved that the physical requirement of diffeomorphism invariance (more precisely, the unitary implementation of the action of the $\textit{Diff}$ group) selects a unique representation of the quantum algebra. This important result established the uniqueness of the kinematics of LQG. 
In loop quantum cosmology quantum kinematics is constructed by using spatial homogeneity to simplify the holonomy-flux algebra,
obtaining a much smaller reduced algebra \cite{abl2003}. Ashtekar and Campiglia \cite{ac2012} used the residual diffeomorphisms to select uniquely a representation of the reduced algebra in the case of the homogeneous, anisotropic Bianchi I spacetime. This uniqueness result was extended to the isotropic case in \cite{eh2016, eht2016}. Collectively these results proved for loop quantum cosmology what the authors of \cite{lost2006} did for the full theory: the uniqueness of its kinematical representation.

If we now turn our attention to dynamics, the situation is markedly different. It has been notoriously difficult to define the Hamiltonian constraint operator in the full theory. The primary reason is that, while the classical constraints are written in terms of local fields such as the connection and the curvature, the variables generating the quantum algebra such as the holonomies are manifestly non-local. Classically one can obtain the local fields from the limiting behavior of a set of holonomies around a loop as one shrinks the loop to a point.
In the seminal work \cite{thiemann1996, thiemann1998, thiemann2007}, 
Thiemann was able to give a prescription for a well-regulated quantization of the Hamiltonian constraint.
However, upon removing the regulator, the resulting operator depends on certain discrete structures chosen in the 
regularization.  
As a consequence the final \textit{physical} space of solutions depends upon the choice of such structures. More recently it has been proposed by Laddha and Varadarajan \cite{tv2012, hlt2012, laddha2014} to constrain such choices using the requirement of four dimensional diffeomorphism covariance by imposing an anomaly-free representation of the Poisson bracket algebra \cite{hkt1976}. So far, however, a definition of the Hamiltonian constraint independent of arbitrary choices remains an open issue in LQG.   

In the simplified setting of loop quantum cosmology the quantum Hamiltonian constraint has also been formulated for many models, including in the isotropic case by Ashtekar, Pawlowski and Singh (APS) \cite{aps2006} and the Bianchi I case by Ashtekar and Wilson-Ewing (AW) \cite{aw2009a}. However, as in the full theory, one has to make arbitrary choices to obtain these results. Naturally one is led to ask whether physical requirements can restrict such ambiguities, as has already been shown for quantum kinematics, at least in the simplified cosmological context.

In the present work we provide a positive answer to this question. Starting with a very general ansatz for a quantum operator in the Bianchi I model and imposing the invariance under residual diffeomorphisms, we arrive at a highly restricted set of possibilities for the quantum Hamiltonian constraint. By requiring that the operator has as its classical limit the classical Hamiltonian constraint of Bianchi I, as well as imposing a certain minimality principle and `planar loops' condition, we obtain the Hamiltonian constraint operator of AW \cite{aw2009a}. Furthermore, by using the AW projector we can restrict to the isotropic model and obtain, notably \textit{without} any planar loops condition, the Hamiltonian constraint first written down by APS \cite{aps2006}.

The present work is not the first to investigate how diffeomorphism invariance constrains ambiguities in 
LQC dynamics. Prior work by Corichi and Singh \cite{cs2008a} investigated how invariance under 
passive diffeomorphisms (namely, rescaling of the background structure --- the fiducial cell)
constrains ambiguities in isotropic LQC. In that work, the authors started from
a one-parameter family of possible quantizations and selected exactly one of them by imposing such diffeomorphism invariance on the resulting physical predictions. The aim of the present work is much broader, imposing invariance on the full quantum Hamiltonian operator, and starting not from a one-parameter family, but from the set of all possible operators in Bianchi I LQC. We also take the active, rather than the passive, view of diffeomorphisms, though these are equivalent.

This paper is structured as follows. In \ref{sec:prelim} we review the classical Bianchi I model and the kinematics of Bianchi I LQC, and  in \ref{sec:strategy} we outline our overall strategy. In \ref{sec:BIham} we implement our strategy and derive the quantum Hamiltonian of Bianchi I LQC. In \ref{sec:proj} we project the Bianchi I Hamiltonian to the isotropic model and obtain the LQC Hamiltonian for $k=0$ Friedmann-Lema\^{i}tre-Robertson-Walker (FLRW) cosmology. We close in \ref{sec:disc} by discussing the results.

\section{Preliminaries}
\label{sec:prelim}

\subsection{Review of Bianchi I LQC}
In this section we briefly review the classical dynamics of the Bianchi I model (for more details, see\cite{aw2009a}). The Bianchi I spacetime is the simplest homogeneous, anisotropic solution of Einstein's equations characterized by the diagonal line element:
\begin{displaymath}
ds^2 = -N^2(t) dt^2 + a^2_1(t) dx^2_1 + a^2_2(t) dx^2_2 + a^2_3(t) dx^2_3,
\end{displaymath}
where $a_1(t), a_2(t), a_3(t)$ are the independent directional scale factors and $N(t)$ is the lapse function.

The action of the symmetry group (the translation group) provides left-invariant $1$-forms $\fom^i_a$ (fiducial co-triads) and left-invariant vector fields $\fe^a_i$ (fiducial triads). The fiducial co-triads give the fiducial metric $\fq_{ab} = \fom^i_a \fom^j_a \delta_{ij}$, the determinant of which we denote by $\fq$. Because the fields are homogeneous on the non-compact slice, writing down the Hamiltonian requires introducing an infrared regulator for the integrals. Such a regulator is furnished by using a fiducial cell $\mathcal{V}$ adapted to the fiducial triads with the lengths of the three cell edges $L_1, L_2, L_3$ and the volume $V_o = L_1 L_2 L_3$ measured with respect to the fiducial metric $\fq_{ab}$. The physical triads are $e^a_i = a^{-1}_i \fe^a_i$, and the physical co-triads are given by $\omega^i_a = a^i \fom^i_a$, so that the physical metric is $q_{ab} = \omega^i_a \omega^j_a \delta_{ij}$ with the determinant $q$.

The basic variables in LQG are the $\SU$ connection $A^i_a$ and the densitized triad $E^a_i$. By fixing the gauge they can be written as
\beq
\label{eq:defcp}
A^i_a = c^i (L^i)^{-1} \fom^i_a \, , \quad\quad\quad E^a_i = p_i L_i V^{-1}_o \sqrt{\fq}\fe^a_i \, ,
\eeq
with $c^i, p_i$ constants. Therefore, the phase space is six-dimensional and parametrized by $c^i, p_i$. The non-vanishing Poisson bracket is given by 
\beq
\label{eq:pbra}
\{ c^i, p_j \} = 8\pi G \gamma \delta^i_j,
\eeq
where $G$ is the Newton constant and $\gamma$ the Barbero-Immirzi parameter. 

The only non-trivial constraint that has to be imposed is the Hamiltonian constraint. It is given by integrating the Hamiltonian density over the fiducial cell:
\begin{displaymath}
C_H = \int_{\mathcal{V}} N \mathcal{H}\, \dd^3 x,
\end{displaymath}
where the Hamiltonian density $\mathcal{H}$ is 
\begin{displaymath}
\mathcal{H} = \frac{E^a_i E^b_j}{16 \pi G \sqrt{|q|}} (\epsilon^{ij}_{\;\;k} F^k_{ab} - 2(1+\gamma^2)e^{ci}e^{dj}K_{c[a} K_{b]d}).
\end{displaymath}
Here $F^k_{ab}$ is the curvature of the connection $A^i_a$ and $K_{ab}$ is the extrinsic curvature. 

Because of the Bianchi I symmetry the Hamiltonian density can be simplified and written as
\begin{displaymath}
\mathcal{H} = -\frac{\sqrt{\fq}}{8\pi G \gamma^2 \sqrt{|p_1 p_2 p_3|} V_o} (p_1 p_2 c_1 c_2 + p_1 p_3 c_1 c_3 + p_2 p_3 c_2 c_3).
\end{displaymath}
We assume the lapse $N$ to be a function of the volume $v:=\sqrt{|p_1 p_2 p_3|}$ only, with the form $N(v)=v^n$ for $n$ a real number.
This in particular (for $n=1$) includes the choice of the lapse $N = \sqrt{|p_1 p_2 p_3|}=v$ used in \cite{aw2009a}.
Integrating over the fiducial cell we then obtain the constraint
\beq
\label{eq:hamc}
C_H  = -\frac{1}{8\pi G \gamma^2} v^{n-1} (p_1 p_2 c_1 c_2 + p_1 p_3 c_1 c_3 + p_2 p_3 c_2 c_3).
\eeq
From now on, this is the expression we will refer to as the classical Hamiltonian constraint of the Bianchi I model.

\subsection{Residual diffeomorphism symmetries}
In the previous section we fixed the gauge by choosing fiducial background structures and 
requiring $A^i_a, E^a_i$ to have the form \eqref{eq:defcp}. However, this gauge-fixing does not remove diffeomorphism freedom completely. The remaining freedom is referred to as the group of \textit{residual diffeomorphisms}. There is a three-parameter family of diffeomorphisms that preserve the form \eqref{eq:defcp} and have a non-trivial action on $c^i, p_j$. These are anisotropic dilations: $x_1 \mapsto e^{\lambda_1}x_1, x_2 \mapsto e^{\lambda_2}x_2, x_3 \mapsto e^{\lambda_3}x_3$. Under their action $c^i, p_j$ transform as
\begin{align}
\label{eq:classdil}
\vec{c} \mapsto \Lambda_c(\vec{\lambda}) \vec{c}, \qquad &\vec{p} \mapsto \Lambda_p(\vec{\lambda}) \vec{p},\\
\label{eq:LcLpdef}
\text{with}\quad \Lambda_c(\vec{\lambda}) := \mathrm{diag}(e^{\lambda_1}, e^{\lambda_2}, e^{\lambda_3}), 
\quad &\Lambda_p(\vec{\lambda}) := \mathrm{diag}(e^{\lambda_2 + \lambda_3}, e^{\lambda_3 + \lambda_1}, e^{\lambda_1 +\lambda_2}).
\end{align}
To be a canonical symmetry a transformation has to preserve the Poisson bracket structure \eqref{eq:pbra}. It is easy to see that this requires the action of dilations to be volume-preserving, $\lambda_1 + \lambda_2 + \lambda_3 = 0$. The $2$-dimensional group of volume-preserving dilations
defines the residual continuous canonical symmetries of the Bianchi I model.
The remaining volume-changing dilations form the \textit{non-canonical} symmetries.

In addition to the residual continuous diffeomorphisms, there are also residual discrete diffeomorphisms. There are three parity transformations $\Pi_1,\Pi_2,\Pi_3$. $\Pi_1$ is the diffeomorphism which, in the preferred coordinates $x_i$, takes the form $(x_1, x_2, x_3) \mapsto (-x_1, x_2, x_3)$, and similarly for $\Pi_2, \Pi_3$. In the present homogeneous context this physical space definition of parity in terms of orientation-changing diffeomorphisms is equivalent to the definition of parity acting in the internal space, where $\Pi_1$ corresponds to the constant $O(3)$ gauge rotation $\mathrm{diag}(-1,1,1)$ which maps $(E^a_1, E^a_2, E^a_3) \mapsto (-E^a_1, E^a_2, E^a_3)$ and $(A^1_a, A^2_a, A^3_a) \mapsto (-A^1_a, A^2_a, A^3_a)$.
The resulting action of the parity operation $\Pi_1$ on the canonical variables $c_i, p_i$ in Bianchi I is given by
\begin{equation}
\label{eq:classpar}
\Pi_1 (c_1, c_2, c_3) = (-c_1, c_2, c_3), \quad\quad \Pi_1 (p_1, p_2, p_3) = (-p_1, p_2, p_3),
\end{equation}
and similarly for the actions of $\Pi_2, \Pi_3$.\footnote{In \cite{aw2009a} the authors took the viewpoint that the Immirzi parameter is an internal pseudoscalar with a dynamically determined sign equal to $\sgn(p_1 p_2 p_3)$,
which leads to an apparently different action of parity.
We do not take this view of the Immirzi parameter because it can not be extended to full LQG. That being said, 
the resulting (Bianchi I) framework is equivalent.}

The rest of the discrete residual diffeomorphisms arise from different combinations of reflections about the $x=y, x=z$ or $y=z$ planes. The action of these symmetries is equivalent to permuting the components of $\vec{c}$ and $\vec{p}$.
We thus label them by permutations $\sigma \in S_3$,
\begin{equation}
\label{eq:classperm} 
\sigma (\vec{c}, \vec{p}) = (\sigma \vec{c}, \sigma \vec{p}). 
\end{equation}

Note that the Hamiltonian constraint \eqref{eq:hamc} is invariant under all of the \textit{canonical} symmetries described above.
Under the non-canonical symmetries --- that is, the volume-changing dilations --- it is \textit{covariant}, with scaling law
$C_H \mapsto e^{(n+1)(\lambda_1 + \lambda_2+\lambda_3)} C_H$.

\subsection{Quantum kinematics}
\label{subsec:prelimquant}

The kinematical space of states in Bianchi I LQC \cite{aw2009a} 
is the Bohr Hilbert space of \textit{almost periodic functions} $\psi(\vec{c})$
on $\mathbb{R}^3$. The basic phase space functions with direct quantum operator analogues are $p_i$ and $e^{i\vec{\mu}\cdot\vec{c}}$ (the second of these classes of functions will be generalized in section \ref{subsec:classan}, following the ideas of \cite{aps2006} and \cite{aw2009a}).
 The eigenstates $|\vec{p}\rangle = |p_1, p_2, p_3\rangle$ of the $\hat{p}_i$ operators form an orthonormal 
basis of this Hilbert space.  A general element of this space is thus of the form 
\begin{displaymath}
|\Psi \rangle = \sum_{\vec{p}} \Psi(\vec{p}) |\vec{p}\rangle ,
\end{displaymath}
with $\Psi(\vec{p})$ non-zero only for a countable number of $\vec{p}$'s and satisfying
\begin{displaymath}
\sum_{\vec{p}} |\Psi(\vec{p})|^2  < \infty.
\end{displaymath}
The action of the basic operators is given by $\hat{p}_i |\vec{p}\rangle = p_i |\vec{p}\rangle$
and $\widehat{e^{i\vec{\mu}\cdot \vec{c}}} |\vec{p}\rangle = |\vec{p} + 8\pi \gamma G \hbar \vec{\mu} \rangle$.
The action of the dilations \eqref{eq:classdil} on quantum states, \textit{in the volume-preserving case}, is given 
by $\Lambda(\l) | \vec{p} \rangle = |\Lambda_p(\l) \vec{p}\rangle$.
The action of \textit{volume-changing} dilations is discussed and defined in section \ref{subsec:noncandiff} of this paper.
The actions of parity $\Pi_l$ \eqref{eq:classpar} and permutations $\sigma$ \eqref{eq:classperm} are given by 
$\Pi_l |\vec{p}\rangle = |\Pi_l \vec{p}\rangle$ and $\sigma |\vec{p}\rangle = |\sigma \vec{p}\rangle$.

\section{Overall strategy}
\label{sec:strategy}

Let us briefly review the overall strategy pursued in this paper. First, we require the quantum Hamiltonian operator to preserve the space of kinematical LQC states, namely the almost-periodic functions on $\mathbb{R}^3$. Then we assume the most general form (of which we are aware) for such an operator that ensures a well-defined classical analogue. Then, after requiring the operator to be hermitian, we impose the \textit{canonical} residual symmetries described above, namely parity invariance, permutation invariance and invariance under the volume-preserving dilations. These requirements already provide significant constraints on the form of the Hamiltonian.

We then establish the action of the \textit{non-canonical} symmetries, i.e., the volume-changing dilations, and impose covariance under them. We require that $\lp:= \sqrt{\hbar G}$ be the only length scale in the structure of the operator and impose that, in the classical limit, the Hamiltonian operator reduce to the classical Hamiltonian constraint. Finally, we use certain simplicity principles. Namely, we constrain the number of terms in the operator to be minimal and require that the curvature be quantized using holonomies around planar loops only. These reasonable simplifications lead to the quantum Hamiltonian having, to both leading and subleading orders in $\lp$, \textit{exactly} the form introduced in \cite{aw2009a} for Bianchi I LQC. 
To address the isotropic case, the planar loops condition is not needed, so we do not impose it. After projecting down to the isotropic model
and requiring minimality, we obtain the APS Hamiltonian \cite{aps2006}.

\subsection*{Remark on the treatment of non-canonical symmetries}

Only the canonical symmetries preserve the symplectic structure of the theory and hence can be implemented in quantum theory
by unitary operators in a way which is consistent with the elementary commutation relations. The non-canonical symmetries --- that is, the volume-changing dilations --- must be treated in a distinct way - we provide in section \ref{subsec:noncandiff} a definition of their action directly on operators not arising from the action of any unitary operator on states.

The reader may wonder why we do not use the same technique as that used in the recent work \cite{eht2016}
on uniqueness of \textit{kinematics} of LQC to make the volume-changing dilations canonical, and hence avoid this complication. The technique used in \cite{eht2016} involved two steps: first let the dilations act on the fiducial cell, so that $V_o$ is also rescaled by the action of dilations.  
This first step directly addressed the heart of the problem: that a background structure, the fiducial cell, had been introduced, breaking the volume-changing part of the residual diffeomorphism symmetry.  To recover volume-changing dilations as a symmetry of the theory, one simply needed to let them act on this background structure. However, this leads to another problem: the operator in the quantum theory corresponding to $V_o$ 
is a multiple of the identity and so will never scale with any potential action of dilations via linear transformations on the state space. 
This latter problem was solved by the second step: to use the \textit{momentum $\pi$ strictly canonical conjugate} to $c$, so that the Poisson brackets are $1$, and $V_o$ \textit{is entirely removed from the framework required for the question}. 

This same strategy \textit{no longer works} for the purposes of the present paper, since once one replaces $p$ in favor of $\pi$, $V_o$ appears in other
expressions, in particular the Hamiltonian constraint. The strategy only works for questions dealing with kinematics, not dynamics.

\section{Selection of a Bianchi I Hamiltonian constraint from physical assumptions}
\label{sec:BIham}
\subsection{Preservation of the space of almost periodic functions on \texorpdfstring{$\mathbb{R}^3$}{ℝ\textthreesuperior}}

The basic Poisson algebra at the root of loop quantum gravity is that generated by holonomies and fluxes.  
In order for the restriction of this Poisson algebra to the Bianchi I phase space to again be closed under Poisson brackets, 
it is necessary to consider instead the subalgebra in which only holonomies along the three symmetry axes are included 
\cite{eht2016}.  The works \cite{eh2016, eht2016} prove that the analogous assumption in the isotropic model does not affect the final quantum framework, suggesting that here likewise it has no such effect. 
The representations of the universal enveloping algebra generated by this subalgebra 
coincide with representations of the \textit{Weyl algebra}, generated by $\widehat{\exp(i\vec{\mu}\cdot\vec{c})}$ and $\widehat{\exp(i\vec{\eta}\cdot\vec{p})}$, which are continuous in the coefficients $\eta^i$ of the $p_i$'s.
The latter were investigated by Ashtekar and Campiglia \cite{ac2012}, who found that there is only one such 
representation that is cyclic and in which residual diffeomorphisms are unitarily implemented.
This unique representation is in fact the standard one used in Bianchi I LQC \cite{aw2009a}, in which the Hilbert space of states is the space of \textit{almost periodic functions}
on $\mathbb{R}^3$.  

The use of the almost periodic functions on $\mathbb{R}^3$ as our Hilbert space of states is thus 
strongly determined by physical principle.
In this paper we 
turn to dynamics; but the first requirement that we impose for our Hamiltonian constraint operator is that it 
\textit{preserve this Hilbert space}.  Any operator $\hat{H}$ satisfying this requirement will map each eigenstate of momentum $\p$
into a countable linear combination of eigenstates of momentum, and hence will have an action taking the form
\begin{align}
\label{presformstate}
\hat{H} | \p\rangle = \sum^{\tilde{N}}_{i=1} g_i(\p) |\F_i(\p)\rangle
\end{align}
with $\tilde{N}$ possibly infinite, for some set of maps $g_i: \mathbb{R}^3 \rightarrow \mathbb{C}$
and $F_i: \mathbb{R}^3 \rightarrow \mathbb{R}^3$.
For each $F: \mathbb{R}^3 \rightarrow \mathbb{R}^3$ define the translation operator
\begin{align*}
T_F |\p\rangle := |\F(\p)\rangle .
\end{align*}
Then $\hat{H}$ takes the form
\begin{align}
\label{presform}
\hat{H} = \sum^{\tilde{N}}_{i=1} T_{\!F_i}\, g_i(\vec{p}).
\end{align}

\subsection{Hermiticity}

We require $\hat{H}$ to be hermitian. This implies that for each term $T_{\!F_i}\, g_i(\p)$ in \eqref{presform}, among the rest of the
terms must exist $\overline{g_i(\p)} \, T_{F_i}^\dagger$. Now
\begin{align*}
T_F^\dagger  |\p\rangle = \sum_{\pvec{p}' \in F^{-1}(\{\p\})} |\pvec{p}'\rangle ,
\end{align*}
so that, when acting on any eigenstate $| \pvec{p}'\rangle$,
\begin{align*}
\overline{g_i(\p)} \, T_{F_i}^\dagger |\pvec{p}'\rangle = \overline{g_i(\p)} \sum_{\pvec{p}'' \in F_i^{-1}(\{\pvec{p}'\})} |\pvec{p}''\rangle
= \sum_{\pvec{p}'' \in F_i^{-1}(\{\pvec{p}'\})} \overline{g_i(\pvec{p}'')}  |\pvec{p}''\rangle
\end{align*}
which again fits into the form (\ref{presformstate}) as expected.  However, to make the hermiticity of 
$\hat{H}$ manifest, from now on we write it simply as
\begin{align}
\label{saform}
\hat{H} = \sum^N_{i=1} \left(T_{\!F_i}\, g_i(\p) + \overline{g_i(\p)} \, T_{\!F_i}^\dagger \right)
= \sum^N_{i=1} \left(T_{\!F_i}\, g_i(\p)  + \hc\right) ,
\end{align}
where ``h.c.'' stands for hermitian conjugate.

\subsection{Existence of a classical analogue}
\label{subsec:classan}

We now make an assumption about the $F_i$'s:

\textbf{Assumption 1.} Each $F_i: \mathbb{R}^3 \rightarrow \mathbb{R}^3$ is generated as the flow, evaluated at unit time, of some vector field $8\pi \gamma G\hbar \f_i(\p)$ on $\mathbb{R}^3$.

The reason for this assumption is as follows.  Each term in $\hat{H}$ corresponds to a translation operator with a coefficient. 
Only if the above assumption is satisfied can each shift operator be cast as the quantization of an exponential, 
so that $\hat{H}$ takes the form
\beq
\label{qH}
\hat{H} = \sum^N_{i=1} \left(\widehat{e^{i \f_i(\p) \cdot \vec{c}}}\, g_i(\p) 
+ \overline{g_i(\p)} \left(\widehat{e^{i \f_i(\p) \cdot \vec{c}}}\right)^\dagger \right).
\eeq
When cast in this form, $\hat{H}$ has an immediate classical phase space function analogue.
This analogue is then central to the state-independent way of taking the classical limit which we use in this paper.
By contrast, if the above assumption is not satisfied, we are not aware of any state-independent way to associate with it a classical phase space function.

Before continuing, we remark on the case in which a map $F_i : \mathbb{R}^3 \rightarrow \mathbb{R}^3$
is \textit{not onto} $\mathbb{R}^3$. The condition that $F_i$ be generated by a vector field does not preclude this 
possibility. A very simple vector field $\vec{f}(\vec{p})$ which shows this is
\begin{align*}
\f(\vec{p}) := (\sgn \,p_x,  0, 0),
\end{align*}
where for definiteness in the above expression we define $\sgn(0) = 0$.
The flow generated by $8\pi \gamma G\hbar \vec{f}(\vec{p})$ is then
\begin{align*}
F(\p) = \left\{ \begin{array}{cc} (p_x+1, p_y, p_z) & \text{if } p_x > 0 \\
(0, p_y, p_z) & \text{if } p_x = 0 \\
(p_x - 1, p_y, p_z) & \text{if } p_x < 0 \end{array} \right. ,
\end{align*}
which has as its image $((-\infty,-1) \cup \{0\} \cup (1, \infty)) \times \mathbb{R} \times \mathbb{R}$, and so is not onto $\mathbb{R}^3$.
$T_F$ has as its interpretation the quantization
of the classical quantity $e^{i\vec{f}(\p) \cdot \vec{c}}$.
One calculates the hermitian conjugate of $T_F$ to be
\begin{align*}
T_F^\dagger |p_x, p_y, p_z \rangle
= \left\{ \begin{array}{cc} |p_x - 1, p_y, p_z \rangle & \text{if } p_x > 1 \\
|0, p_y, p_z \rangle & \text{if } p_x = 0 \\
|p_x + 1, p_y, p_z \rangle & \text{if } p_x < -1 \\
0 & \text{if } p_x \in (-1,0)\cup(0,1)
\end{array} \right. .
\end{align*}
Note that $T_F^\dagger$ has a kernel and that this is a direct consequence of the fact that $F$ is not onto.
One of the standard quantization axioms is that hermitian conjugates in quantum theory should correspond to complex conjugation of the
corresponding classical quantities. We would therefore like the quantization rules to be defined in such a way that 
$T_F^\dagger$ is the quantization of $e^{-i\vec{f}(\p) \cdot \vec{c}}$.
In fact, the flow generated by $-8\pi \gamma G\hbar \vec{f}(\vec{p})$ is
\begin{align*}
\tilde{F}(\p) = \left\{ \begin{array}{cl}(p_x - 1, p_y, p_z ) & \text{if } p_x > 1 \\
(0, p_y, p_z ) & \text{if } p_x = 0 \\
(p_x + 1, p_y, p_z ) & \text{if } p_x < 1 \\
\text{not defined} & \text{if } p_x \in (-1,0)\cup(0,1) \end{array} \right. .
\end{align*}
This motivates us to extend the definition of $T_F$ to the case where $F$ is not defined on all of $\mathbb{R}^3$:
\begin{align}
\label{TFext}
T_F |\p\rangle := \left\{ \begin{array}{cc} |\vec{F}(\p)\rangle & \text{if } \p \in {\rm Dom}(F) \\
0 & \text{if } \p \notin {\rm Dom}(F)\end{array}\right. .
\end{align}
With this definition, if we quantize $e^{-i\vec{f}(\p) \cdot \vec{c}}$ as $T_{\tilde{F}}$, then 
\begin{equation}
\label{starhom}
\left(\widehat{e^{i\vec{f}(\p) \cdot \vec{c}}}\right)^\dagger = \widehat{e^{-i\vec{f}(\p) \cdot \vec{c}}}
\end{equation}
as desired. 

\textit{Thus, to ensure (\ref{starhom}), we make the following general quantization rule:}
When the flow $F(\p)$ generated by a vector field $8\pi \gamma G\hbar \f(\p)$ is not globally defined on $\mathbb{R}^3$,
then we quantize it as $T_F$ with $T_F$ defined as in (\ref{TFext}). 
The form (\ref{qH}) now becomes
\begin{equation}
\label{qHcomplete}
\hat{H} = \sum^N_{i=1} \left(\widehat{e^{i \f_i(\p) \cdot \vec{c}}}\,\, g_i(\p) + 
\overline{g_i(\p)} \,\,\widehat{e^{-i \f_i(\p) \cdot \vec{c}}}\right).
\end{equation}

We have seen above that assumption 1 does not preclude the possibility that an $F_i$ not be onto or even be 
ill-defined for some arguments.  However, as an aside, we wish to point out one general restriction on the $F_i$'s which \textit{is}
imposed by assumption 1, just to show that the assumption is non-trivial. According to assumption 1, $F_i$ arises as the $t=1$
evaluation of the flow $F_i^t$ generated by $8\pi \BI G\hbar \f_i$.  By taking the gradient of the flow equation defining $F_i^t$, taking the 
determinant of both sides, and solving the resulting ordinary differential equation in $t$, one obtains
\begin{align}
\label{detsol}
\det \left( \frac{\partial \vec{F}_i^t}{\partial \pvec{p}'}\right)
= \exp \left( \int_0^t \det \left(\frac{\partial \vec{f}_i}{\partial \vec{p}}\left(F_i^{t'}\left(\pvec{p}'\right) \right)dt'\right)
\right),
\end{align}
which in particular implies 
\begin{align}
\label{posdet}
\det \left(\frac{\partial \vec{F}^t_i}{\partial\pvec{p}'} \right) > 0
\end{align}
\textit{always}, so that $F_i$ is required to be orientation-preserving wherever it is defined.
This excludes $F_i$ from being, for example,
the parity map. 

The form (\ref{qHcomplete}) for $\hat{H}$ has a direct classical analogue; however, the classical analogue is not yet uniquely defined, 
because there is always more than one way to cast $\hat{H}$ into the form (\ref{qHcomplete}). This can be seen from the fact that
\begin{align}
\label{eq:classamb}
g(\p)\,\, \widehat{e^{i\f(\p)\cdot \c}} = \widehat{e^{i\f(\p)\cdot \c}}\,\, g(\F(\p)).
\end{align}
For the moment, we leave this ambiguity free, fixing it later in subsection \ref{subsec:fixclassan}.
The one thing we do require at this point (without loss of generality) is that, using the above identity, all terms with the same $\f_i(\p)$ 
in (\ref{qHcomplete}) be combined, so that (\ref{qHcomplete}) has the property
\begin{align}
\label{eq:uniqf}
\vec{f}_i = \pm \vec{f}_j \; \text{implies} \; i=j.
\end{align}
This property will be important in the proofs below.

In light of the above discussion of the issue of the classical analogue of $\hat{H}$, we will, for convenience,
in general not write hats over the quantization of the exponentials in (\ref{qHcomplete}), except when needed for clarity.

\subsection{Invariance under canonical residual diffeomorphisms}

We now impose invariance under residual diffeomorphisms which are also canonical transformations.

\subsubsection{Volume-preserving dilations}

We begin by requiring that $\hat{H}$ be invariant under the continuous canonical residual diffeomorphisms, namely volume-preserving dilations $\Lambda(\l)$ with $\lambda_1 + \lambda_2 + \lambda_3 = 0$. 
To do this, we first clarify the action
of these dilations on the operators involved in $\hat{H}$.
It is easy to check that, for any invertible linear mapping $L:\mathbb{R}^3 \rightarrow \mathbb{R}^3$,
if $F: \mathbb{R}^3 \rightarrow \mathbb{R}^3$ is the unit-time flow generated by 
$8\pi \gamma G \hbar f: \mathbb{R}^3 \rightarrow \mathbb{R}^3$, then $(L^{-1}\circ F \circ L)$ is the 
unit-time flow generated by $8\pi \gamma G \hbar (L^{-1} \circ f \circ L)$.
That is, the association between $f$ and $F$ is covariant with respect to the adjoint action of linear maps on $\mathbb{R}^3$.
This is true in particular for the case $L = \Lambda_p(\l)$ \eqref{eq:LcLpdef}.
From this one can deduce that the action of $\Lambda(\l)$ on the operators $\hat{p}_i$ and the exponentials \eqref{starhom} is given by
\begin{align}
\begin{split}
\Lambda(-\l) \hat{p}^i \Lambda(\l) &= e^{-\lambda_i} p^i, \\
\Lambda(-\l) \widehat{e^{i\vec{f}(\vec{p})\cdot \vec{c}}} \Lambda(\l)
&= \reallywidehat{e^{i(\Lambda_p(-\l)\vec{f}(\Lambda_p(\l)\vec{p}))\cdot \vec{c}}}
= \reallywidehat{e^{i\vec{f}(\Lambda_p(\l)\vec{p})\cdot (\Lambda_c(\l)\vec{c})}}.
\end{split}
\label{eq:candilaction}
\end{align}
The invariance requirement then yields
\begin{align*}
\Lambda(-\l) \hat{H} \Lambda(\l) &= \Lambda(-\l) \sum^N_{i=1} \left(e^{i \f_i(\p) \cdot \vec{c}} g_i(\p) + \overline{g_i(\p)} e^{-i \f_i(\p) \cdot \vec{c}}\right) \Lambda(\l) \\ &= \sum^N_{i=1} \left(\Lambda(-\l)  e^{i \f_i(\p) \cdot \vec{c}} \Lambda(\l) \Lambda(-\l) g_i(\p) \Lambda(\l) + \Lambda(-\l) \overline{g_i(\p)} \Lambda(\l) \Lambda(-\l) e^{-i \f_i(\p) \cdot \vec{c}} \Lambda(\l)\right) \\ &= \sum^N_{j=1} \left(e^{i \f_j(\p) \cdot \vec{c}} g_j(\p) + \overline{g_j(\p)} e^{-i \f_j(\p) \cdot \vec{c}}\right)  = \hat{H},
\end{align*}
which implies that for any term labeled by $i=1,2,\dots,N$ there exists a term labeled by $j=1,2,\dots,N$ such that
\beq
\label{eq:vpdiffg}
\Lambda(-\l) g_i(\p) \Lambda(\l) = g_i(e^{-\lambda_1} p_1, e^{-\lambda_2} p_2, e^{\lambda_1+\lambda_2} p_3) = g_j(p_1, p_2, p_3) = g_j(\p)
\eeq
and 
\bq
\Lambda(-\l) e^{i \f_i(\p) \cdot \vec{c}} \Lambda(\l) = e^{i \f_j(\p) \cdot \vec{c}},
\eq
which in turn implies 
\beq
\label{eq:vpdiff}
e^{\lambda_k} f_i^k(e^{-\lambda_1} p_1, e^{-\lambda_2} p_2, e^{\lambda_1+\lambda_2} p_3) = f_j^k(p_1, p_2, p_3)
\eeq
for $k=1,2,3$.

Then by taking $\l=\vec{0}$ in \eqref{eq:vpdiff} and using \eqref{eq:uniqf} we obtain that $i=j$ in \eqref{eq:vpdiffg} and \eqref{eq:vpdiff}, and therefore
\beq
\label{eq:vpdiff2}
e^{\lambda_k} f_i^k(e^{-\lambda_1} p_1, e^{-\lambda_2} p_2, e^{\lambda_1+\lambda_2} p_3) = f_i^k(\p) \quad\quad  g_i(e^{-\lambda_1} p_1, e^{-\lambda_2} p_2, e^{\lambda_1+\lambda_2} p_3) =  g_i(\p).
\eeq

Because the diffeomorphisms under consideration preserve volume, we are led to rewrite these equations using as the third variable the volume $v$ (defined above as $v=\sqrt{|p_1 p_2 p_3|}$ and assumed here to be $v\neq 0$) instead of $p_3$ and obtain
\begin{align}
\label{eq:vpdiff3}
e^{\lambda_k} f_i^k(e^{-\lambda_1} |p_1|, e^{-\lambda_2} |p_2|, v, \sgn p_1, \sgn p_2, \sgn p_3) &= f_i^k(|p_1|, |p_2|, v, \sgn p_1, \sgn p_2, \sgn p_3) \\  g_i(e^{-\lambda_1} |p_1|, e^{-\lambda_2} |p_2|,  v, \sgn p_1, \sgn p_2, \sgn p_3) &=  g_i(|p_1|, |p_2|, v, \sgn p_1, \sgn p_2, \sgn p_3).
\end{align}

Then, by taking $e^{-\lambda_1} |p_1|=e^{-\lambda_2} |p_2|=1$ we get
\begin{align*}
|p_{(1,2)}| f_i^{(1,2)}(1, 1, v, \sgn p_1, \sgn p_2, \sgn p_3) &= f_i^{(1,2)}(|p_1|, |p_2|, v, \sgn p_1, \sgn p_2, \sgn p_3) \\ \frac{|p_3|}{v^2} f_i^3(1, 1, v, \sgn p_1, \sgn p_2, \sgn p_3) &= f_i^3(|p_1|, |p_2|, v, \sgn p_1, \sgn p_2, \sgn p_3) \\
g_i(1, 1,  v, \sgn p_1, \sgn p_2, \sgn p_3) &=  g_i(|p_1|, |p_2|, v, \sgn p_1, \sgn p_2, \sgn p_3).
\end{align*}

Therefore, by suitable redefinitions of $f_i^k$ and $g_i^k$ we can restrict their dependence on $\p$ to have the following form
\beq
f_i^k(\p) = p^k \tilde{f}^k_i(v, \sgnp), \quad\quad g_i(\p) = g_i(v,\sgnp),
\eeq
where we introduced notation $\sgnp = (\sgn p_1, \sgn p_2, \sgn p_3)$ for brevity.

To summarize the results of this section, we obtained that the requirement of invariance under volume-preserving dilations ensures that the Hamiltonian constraint is given by
\beq
\label{eq:Hvpdiff}
\hat{H} = \sum^N_{i=1} \left(e^{i \sum_k \tf^k_i(v, \sgnp) p^k c_k} g_i(v,\sgnp) + \hc \right).
\eeq

\subsubsection{Parity}
\label{subsubsec:parity}

Now we turn to residual discrete diffeomorphisms. We impose that $\hat{H}$ should be invariant under each of the three parity transformations $\Pi_1, \Pi_2, \Pi_3$. We shall now demonstrate that this requirement implies that functions $\tf_i^k$ and $g_i$ can be taken to be independent of $\sgnp$. 

First, we again use the covariance property, with respect to the adjoint action of any invertible linear map $L$ on $\mathbb{R}^3$, of the association between a vector field on $\mathbb{R}^3$ and the unit-time flow which it generates, this time for the case 
$L = \Pi_l$. From this one deduces the following action of $\Pi_l$ on the operators $\hat{p}_i$ and the exponentials 
\eqref{starhom}:
\begin{equation}
\Pi_l  \hat{p}_i \Pi_l = \widehat{(\Pi_l p)_i}\, , \quad\quad\quad\quad 
\Pi_l \widehat{e^{i\vec{f}(\vec{p})\cdot \vec{c}}} \Pi_l
= \reallywidehat{e^{i\vec{f}(\Pi_l\vec{p})\cdot (\Pi_l \vec{c})}}.
\label{eq:parityaction}
\end{equation}
Imposing invariance under parity on the expression \eqref{eq:Hvpdiff} for $\hat{H}$, we then get 
\begin{align*}
\Pi_l \hat{H} \Pi_l  &= \sum^N_{i=1} \left(\Pi_l  e^{i \sum_k \tf^k_i(v, \sgnp) p^k c_k} \, \Pi_l \cdot \Pi_l \, g_i(v, \sgnp) \Pi_l + \hc \right) \\ &= \sum^N_{i=1} \left( e^{i \tf^k_i(v, \Pi_l\sgnp) p^k c_k} g_i(v, \Pi_l \sgnp) + \hc \right) \\ &= \sum^N_{j=1} \left(e^{i \tf^k_j(v, \sgnp) p^k c_k} g_j(v,\sgnp) + \hc \right)  = \hat{H}.
\end{align*}

This equation can be satisfied either by using functions $\tf_i^k$ and $g_i$ invariant under parity transformations or by including into the ansatz the relevant extra terms generated by the action of these transformations. If the functions $\tf_i^k$ and $g_i$ are invariant under parity, then it is clear that they are independent of $\sgnp$. If they are not invariant under parity, then for the Hamiltonian constraint to be parity-invariant $\hat{H}$ has to include extra terms generated by the parity transformations. In the latter case one has
\begin{align}
\nonumber
\hat{H} = &e^{i \tf^k_i(v,-\sgn p_1, \sgn p_2, \sgn p_3) p^k c_k} g_i(v,-\sgn p_1, \sgn p_2, \sgn p_3) + \\ 
\nonumber
&+ e^{i \tf^k_i(v,\sgn p_1, \sgn p_2, \sgn p_3) p^k c_k} g_i(v,\sgn p_1, \sgn p_2, \sgn p_3)  + \hc  
+ \text{rest of terms}\\ 
\nonumber
\hat{=} &e^{i \tf^k_i(v,-1, \sgn p_2, \sgn p_3) p^k c_k}g_i(v,-1, \sgn p_2, \sgn p_3)  + \\ 
&+ e^{i \tf^k_i(v,1, \sgn p_2, \sgn p_3) p^k c_k}g_i(v,1, \sgn p_2, \sgn p_3) + \hc 
+ \text{rest of terms}.
\label{eq:parident}
\end{align}
Here $\hat{=}$ means equality of operators when acting on eigenstates $|\p \rangle$ such that
\bq
v = |p_1 p_2 p_3|^{\frac{1}{2}} > v_0 := \sup_{\substack{\alpha, \\ \p \,\, \mathrm{s.t.}\, v=0}} \Big|\prod_k F_{\alpha}(\p)^k \Big|^{\frac{1}{2}},
\eq
where $\alpha$ ranges over all terms in \eqref{eq:parident}, including hermitian conjugates, and $F_{\alpha}: \R^3 \rightarrow \R^3$ is the flow map for the shift operator in the corresponding term. In section \ref{sec:proj} $v_0$ will be calculated, and will turn out to 
be\footnote{If $v=|p_1 p_2 p_3|^{\frac{1}{2}} = 0$ is chosen to be included in the superselected lattice \cite{aps2006, aw2009a}, $v_0$ is equal to the \underline{first} lattice point away from $v=0$.}
on the order of $\lp^3$.
Since the bounce predicted by LQC happens at a value of $v$ at least three orders of magnitude larger than this value \cite{as2011a, aps2006},
the deviation from exact equality allowed by $\hat{=}$ above is completely irrelevant for observational predictions. 
(For the specific Hamiltonian constraint isolated at the end of this paper, when restricted to the resulting superselected lattice,
the above identity (\ref{eq:parident}) even becomes an exact equality.) All operator equalities involving $\hat{H}$ from now on will be understood to use $\hat{=}$.

Equation (\ref{eq:parident}) means that $\tf_i^k$ and $g_i$ can be taken not to depend on $\sgn p_1$. Similarly one can see that $\tf_i^k$ and $g_i$ can be made independent of $\sgn p_2$ and $\sgn p_3$. 
Therefore, the invariance under parity implies that the Hamiltonian constraint can be written as
\beq
\label{eq:Hparity}
\hat{H} \hat{=} \sum^N_{i=1} \left(e^{i \sum_k \tf_i^k(v) p^k c_k} g_i(v) + \hc \right).
\eeq

\subsubsection{Other reflections}

The rest of the discrete residual diffeomorphisms are given by reflections about the $x=y, x=z$ or $y=z$ planes and combinations thereof, whose action on the variables $c_i, p^i$ is equivalent to permuting their components with some element $\sigma$ of the permutation group $S_3$.  The action of such permutations on states was given in \ref{subsec:prelimquant}. 
Using again the covariance property, with respect to the adjoint action of any linear map $L$ on $\mathbb{R}^3$, of the association between $\vec{f}(\p)$ and $\widehat{e^{i \vec{f}(\p)\cdot\vec{c}}}$,
for $L = \sigma$, we obtain the following action of $\sigma$ on the operators $\hat{p}_i$ and the exponentials 
in \eqref{starhom}:
\begin{equation}
\sigma^{-1} \widehat{p_i} \sigma = \widehat{(\sigma p)_i} \, , \qquad \qquad 
\sigma^{-1} \, \widehat{e^{i\f(\p)\cdot \c}}\, \sigma 
=  \reallywidehat{e^{i(\sigma^{-1} \f(\sigma\p))\cdot \c}} =  \reallywidehat{e^{i\f(\sigma\p)\cdot (\sigma \c)}}.
\label{eq:permaction}
\end{equation} 
Imposing invariance under such permutations on the expression \eqref{eq:Hparity} for $\hat{H}$, we get 
\begin{align*}
&\sum^N_{j=1} \left(e^{i \sum_k \tf^k_j(v) p^k c_k} g_j(v) + \hc \right) = \hat{H} = \sigma^{-1} \hat{H} \sigma \\
= &\sum^N_{i=1} \left(\sigma^{-1} e^{i \sum_k \tf^k_i(v) p^k c_k} \, \sigma \cdot \sigma^{-1} \, g_i(v) \sigma + \hc \right) \\
 = &\sum^N_{i=1} \left(e^{i \sum_k (\sigma^{-1} \tf_i)^k(v) p^k c_k} g_i(v) + \hc \right).
\end{align*}
This equation implies that, for each $i$, there exists $j$ such that
\begin{displaymath}
(\sigma^{-1} \tf_i)^k(v) \hat{=} \tf^k_j(v).
\end{displaymath}
This condition can be satisfied by including into the ansatz the relevant extra terms generated by the action of permutations $\sigma$. 

Therefore, the invariance under discrete diffeomorphisms restricts the form of the Hamiltonian constraint to be
\beq
\label{eq:Hdisc}
\hat{H} = \sum_{i} \sum_{\sigma \in S_3} \left(e^{i \sum_k (\sigma \tf_i)^k(v) p^k c_k} g_i(v) + \hc \right).
\eeq

\subsection{Covariance under non-canonical residual diffeomorphisms}
\label{subsec:noncandiff}

In this subsection we address covariance of the Hamiltonian constraint under the remaining residual diffeomorphisms - namely the volume-changing ones, which are the only residual diffeomorphisms which are non-canonical in the sense that they do not preserve the Poisson brackets in the classical theory and hence neither preserve the basic commutators in the quantum theory. A consequence of this non-preservation of 
the commutation relations is that the action of these diffeomorphisms on operators cannot be represented as conjugation by any unitary operator on states. 
The action can, nevertheless, be defined directly on operators.\footnote{The
action which we define is unconventional, but, we believe, very well motivated, as 
invariance under this action leads directly to invariance of the effective equations under volume-changing dilations,
which is basic to the physical viability of the quantum theory.  Nevertheless, because the action is unconventional, we also
include an alternative derivation of the Hamiltonian constraint in the appendix which avoids its use, using instead
a different assumption in its place which is also well-motivated, but not as fundamental.}
Our proposal for this definition requires that we fix a general prescription for correspondence between quantum operators and 
classical phase space functions.
 In subsection \ref{subsec:classan}, a prescription for correspondence between shift operators
and $\mathrm{U(1)}$-valued phase space functions was fixed, but the correspondence between more general operators and phase space functions
was not fixed. We begin by fixing such a correspondence.

\subsubsection{Fixing of a classical-quantum correspondence}
\label{subsec:fixclassan}

As noted in subsection \ref{subsec:classan}, when an operator is cast in the form (\ref{qHcomplete}), it has an immediate classical 
analogue, but, due to equation (\ref{eq:classamb}), this classical analogue depends on the order of the operators chosen.
This problem of fixing a classical analogue for a given operator --- the problem of ``classicalization'' --- 
is the inverse of the usual problem of quantization going in the other direction. 
The dependence of the quantization map on an ordering choice is well-known 
and is the same as the dependence of the classicalization map on such a choice. 
We choose to fix the following symmetric ordering prescription:
\begin{align}
\label{eq:termquant}
\reallywidehat{g(\p) e^{i\f(\p)\cdot \c}} := \frac{1}{2}\left(g(\p) \widehat{e^{i\f(\p)\cdot \c}} + \widehat{e^{i\f(\p)\cdot \c}} g(\p) \right).
\end{align}
The quantization of sums of terms (\ref{eq:termquant}) is then fixed via
\begin{align}
\label{eq:sumquant}
\reallywidehat{\varphi_1 + \varphi_2} := \widehat{\varphi_1} + \widehat{\varphi_2}.
\end{align}
This prescription has the following advantages:
\begin{enumerate}
\item It intertwines complex conjugation and hermitian conjugation: 
$\widehat{\overline{\varphi(\c,\p)}} = \widehat{\varphi(\c,\p)}^\dagger$.
\item It is covariant with respect to all the canonical symmetries $\Lambda$:
$\Lambda^{-1} \widehat{\varphi(\c,\p)} \Lambda = \reallywidehat{\varphi(\Lambda (\c, \p))}$,
where $\Lambda$ here denotes the action of any volume-preserving dilation, parity map, or permutation.
\item The terms involved in this ordering choice, i.e., on the right hand side of (\ref{eq:termquant}), 
are of the form considered in this paper up until now.
\end{enumerate}
Note that any symmetric ordering would satisfy conditions (1) and (2), and (3) is an advantage only for the presentation 
of this paper.  In general we wish to emphasize that there is \textit{more than one valid ordering choice} here. 
However, though this choice affects the exact phase space function which we associate to 
each term (\ref{eq:termquant}) (and hence to sums of such terms (\ref{eq:sumquant})), 
it only does so at an order subleading by at least 
\begin{align}
\label{eq:fgcomm}
\mathcal{O}\left(\frac{\hbar \left\{g(\vec{p}), e^{i\vec{f}(\vec{p})\cdot \vec{c}}\right\}}
{g(\vec{p}) e^{i\vec{f}(\vec{p})\cdot \vec{c}}}\right) = \mathcal{O}\left(\lp^2 \vec{f}(\vec{p}) \cdot \nabla (\ln g(\p))\right).
\end{align}
As we will see in subsection \ref{subsec:classlim}, the vector fields $\vec{f}_i$, and hence the quantities $\vec{f}_i \cdot \nabla (\ln g_i)$, will all be forced to scale as $\lp$, so that such terms
will be subleading by at least $\mathcal{O}(\lp^3)$ and hence will be at least a full order of $\lp$ subdominant relative to the standard quantum corrections characteristic of LQC (which are $\BO(\lp^2)$), and so turn out to be negligible when one considers phenomenological predictions from the theory.
If we also allow orderings in which the shift operators $\widehat{e^{i\vec{f}(\vec{p})\cdot \vec{c}}}$ are separated into parts 
$\widehat{e^{i\pvec{f}'(\vec{p})\cdot \vec{c}}}$, $\widehat{e^{i\pvec{f}''(\vec{p})\cdot \vec{c}}}$
with $\vec{f}_i = \pvec{f}' + \pvec{f}''$,
this affects the phase space function at an order subleading by at least 
\begin{align}
\mathcal{O}\left(\frac{\hbar \left\{e^{i\pvec{f}'(\vec{p})\cdot \vec{c}}, e^{i\pvec{f}''(\vec{p})\cdot \vec{c}}\right\}}
{e^{i\pvec{f}'(\vec{p})\cdot \vec{c}} e^{i\pvec{f}''(\vec{p})\cdot \vec{c}}}\right) = \mathcal{O}\left(\lp^2 [\pvec{f}'(\vec{p}), \pvec{f}''(\vec{p})]\cdot \c\right),
\end{align}
where $[\pvec{f}'(\vec{p}), \pvec{f}''(\vec{p})]$ denotes the commutator of vector fields on $\mathbb{R}^3$. 
Again, because each $\vec{f}$ will be forced to scale as $\lp$, 
as long as we stipulate that at least one of $\pvec{f}'$, $\pvec{f}''$ is chosen to have the same order in $\lp$ 
as $\vec{f}_i$), this means that such terms will be subleading by 
$\mathcal{O}(\lp^3)$ and hence affect neither the leading nor subleading terms of $\hat{H}$ in an 
$\lp$ expansion, and hence will again have negligible effect on phenomenology.

\subsubsection{Definition of the action of non-canonical dilations}

With the association between operators 
$\widehat{\varphi(\c,\p)}$ and classical phase space functions $\varphi(\c,\p)$ fixed, the action of non-canonical
dilations can be defined simply by 
\begin{align}
\Lambda(\vec{\lambda}) \triangleright \widehat{\varphi(\c,\p)} 
:= \reallywidehat{\varphi(\Lambda(\vec{\lambda})(\c,\p))} 
= \reallywidehat{\varphi(\Lambda_c(\vec{\lambda})\c,\Lambda_p(\vec{\lambda})\p)} .
\label{eq:noncanaction}
\end{align}
Note that, due to equations (\ref{eq:candilaction}),(\ref{eq:parityaction}), (\ref{eq:permaction}), this action of non-canonical dilations is a strict generalization of the action
of the canonical residual diffeomorphisms reviewed in section \ref{subsec:prelimquant}. 
For non-canonical residual diffeomorphisms, because
the commutation relations are not preserved, the above definition of the action \textit{depends on the ordering convention} 
(\ref{eq:termquant}) we have used to define the relation between operators and phase space functions. Nevertheless, as noted above, the ambiguity resulting from this choice of ordering only affects the right hand side of (\ref{eq:noncanaction}) to an order which, as we will see, for the Hamiltonian constraint selected, affects neither the dominant nor subdominant contributions to the dynamics in an $\lp$ expansion. 

\subsubsection{Imposing covariance of \texorpdfstring{$\hat{H}$}{Ĥ}}

At this point we are ready to impose that $\hat{H}$ be covariant under the non-canonical dilations.
However, in order to even ask the question whether $\hat{H}$ is invariant under this action, $\hat{H}$ must
first be \textit{in the domain of this action}, that is, it must be in the image of the quantization map defined in subsection
\ref{subsec:fixclassan}. In general $\hat{H}$ is not exactly in this image, but it will always `almost' be in this image.
More precisely, one can always reorder the terms in expression (\ref{eq:Hdisc}) 
to fit the quantization prescription (\ref{eq:termquant}), thereby generating commutator terms whose order 
was calculated in (\ref{eq:fgcomm}):
\begin{align}
\label{eq:Hreorder}
\hat{H} = \hat{H}' + \sum_i \mathcal{O}\left(\lp^2 \, \vec{f}_i(\p)\cdot \nabla (\ln g_i(\p))\right)
= \hat{H}' + \sum_i \mathcal{O}\left(\lp^2 \, \frac{d\ln g(v)}{d\ln v}\sum_k \tilde{f}_i(v)^k \right),
\end{align}
where we have defined
\begin{align}
\nonumber
\hat{H}' &:= \frac{1}{2} \sum_{i} \sum_{\sigma \in S_3} \left(g_i(v) \reallywidehat{e^{i \sum_k (\sigma \tf_i)^k(v) p^k c_k}}
+ \reallywidehat{e^{i \sum_k (\sigma \tf_i)^k(v) p^k c_k}} g_i(v) + \hc \right) \\
&=\sum_{i} \sum_{\sigma \in S_3}  \left(\reallywidehat{g_i(v) e^{i \sum_k (\sigma \tf_i)^k(v) p^k c_k}} + \hc\right) .
\label{eq:Hprime}
\end{align}
That is, the classical analogue of $\hat{H}'$ is precisely
\begin{align}
H :=   \sum_{i} \sum_{\sigma \in S_3} \left(g_i(v) e^{i \sum_k (\sigma \tf_i)^k(v) p^k c_k} 
+ \cc \right),
\end{align}
where ``\cc'' denotes complex conjugate.  
As noted above, $\frac{d\ln g(v)}{d\ln v}\sum_k \tilde{f}_i(v)^k$ will be forced in the next subsection to scale as $\lp$, so that the 
contribution from the commutator terms in (\ref{eq:Hreorder}) will end up being $\mathcal{O}(\lp^3)$,
affecting neither the dominant not subdominant terms in the dynamics. 
Also, note that, just as the form (\ref{eq:Hdisc}) of $\hat{H}$ satisfies all criteria imposed up until now, 
so does the form (\ref{eq:Hprime}) of $\hat{H}'$.

We impose covariance under non-canonical dilations only for the 
part $\hat{H}'$, `almost' equal to $\hat{H}$, which is in the domain of the action of such dilations.
The full group of dilations can be expressed as the direct product of the volume-preserving dilations and the one-dimensional group of
isotropic dilations $\Lambda(\l) := \Lambda((\lambda/3, \lambda/3, \lambda/3))$. The form
(\ref{eq:Hprime}) is already invariant under volume-preserving dilations, so that it remains only to impose only covariance under isotropic dilations. When acting on the form (\ref{eq:Hprime}), the restriction of the action (\ref{eq:noncanaction}) to isotropic dilations maps $\tilde{f}_i^k(v)$ to $e^{\lambda}\tilde{f}_i^k(e^{\lambda}v)$ and $g_i(v)$ to $g_i(e^{\lambda} v)$.  Now, the classical Hamiltonian
(\ref{eq:hamc}), 
when acted upon by classical isotropic dilations, scales by a factor of $e^{(n+1)\lambda}$.  If we require the quantum Hamiltonian operator to
have this same scaling behavior, it follows
\begin{align}
\tilde{f}_i^k(v) &= \tilde{A}_i^k/v, \qquad \text{ with }\tilde{A}_i^k := \tilde{f}_i^k(1) \in \mathbb{R}, \qquad \text{and} \\
g_i(v) &= \tilde{B}_i v^{n+1}, \qquad \text{ with }\tilde{B}_i := g_i(1) \in \mathbb{C}.
\end{align}
The form of the Hamiltonian constraint (\ref{eq:Hreorder}),(\ref{eq:Hprime}) then reduces to
\beq
\label{eq:Hnoncan}
\hat{H} 
= \sum_{i} \sum_{\sigma \in S_3} \left(\reallywidehat{\tilde{B}_i v^{n+1} e^{\frac{i}{v}\sum_k (\sigma \tilde{A}_i)^k p^k c_k}} + \hc \right) + \sum_i \mathcal{O}\left(\lp^2 \, (n+1) \sum_k \tilde{A}^k_i/v \right).
\eeq

Because the relation between quantum operators and classical phase space functions has now been fixed, from now on, when hats are omitted in an operator expression, 
it is understood that the operator indicated is that determined by the prescription
(\ref{eq:termquant}) and (\ref{eq:sumquant}). We will in general do this unless explicit hats aid in clarity.

\subsection{Correct classical limit and unique length scale}
\label{subsec:classlim}

The Hamiltonian operator is a quantization of the classical Hamiltonian constraint and therefore we require that the Hamiltonian reduces to the classical constraint in the classical limit. To take this limit we first introduce the dependence of 
the coefficients defining $\hat{H}$ on a classicality parameter, namely $\lp = \sqrt{\hbar G}$, to obtain
\beq
\label{eq:ham}
\hat{H} 
= \sum_{i} \sum_{\sigma \in S_3} 
\left(\tilde{B}_i(\lp) v^{n+1} e^{\frac{i}{v}\sum_k (\sigma \tilde{A}_i)^k(\lp) p^k c_k} + \hc \right) + \sum_i \mathcal{O}\left(\lp^2 \, (n+1) \sum_k \tilde{A}^k_i(\lp)/v \right).
\eeq
Now, because $\hat{H}$ is not necessarily in the image of the quantization map which we have 
fixed in the last subsection, its classical analogue is not exactly fixed. However, $\hat{H}'$ \textit{does} have an unambiguous 
classical analogue $H$, and since $\hat{H}$ and $\hat{H}'$ are equal in the classical limit,
$H$ may also be chosen as the classical analogue of $\hat{H}$, and we do so. 
Just as $\hat{H}$ depends on $\lp$, so does $H$:
\begin{equation}
\label{eq:Hclassan}
H = \sum_{i} \sum_{\sigma \in S_3} \left(  \tilde{B}_i(\lp) v^{n+1}  e^{\frac{i}{v} \sum_k (\sigma \tilde{A}_i)^k(\lp) p^k c_k} + \cc \right).
\end{equation}
It is the limit of this quantity, as $\lp \rightarrow 0$, that we require to equal the classical constraint.

\subsubsection{Planck length as the unique length scale}

Now, given that $c_i$ is dimensionless and $p_i$ has the dimension of an area, it is clear that $\tilde{A}^k_i(\lp)$ has the dimension of a length. We now require that the \textit{only length scale in the theory} be the \textit{Planck length} $\lp$. This requirement implies that $\tilde{A}^k_i(\lp) = \lp A^{'k}_i$ for dimensionless coefficients $A^{'k}_i$.
Thus, 
\begin{equation}
H = \sum_i \sum_{\sigma \in S_3} \left(\tilde{B}_i(\lp)  v^{n+1}
e^{i \frac{\lp}{v} \sum_k (\sigma A'_i)^k p_k c_k}  + \cc \right) .
\end{equation}

Dimensional arguments can also be applied to fix the form of $\tilde{B}_i(\lp)$. We note that, since $H$ equals the Hamiltonian constraint in the classical limit, the dimension of $\tilde{B}_i(\lp) v^{n+1}$ should match the dimension of the classical Hamiltonian constraint \eqref{eq:hamc}. Using again the assumption that $\lp$ is the only length scale in the theory, one can see that $ \tilde{B}_i(\lp) = \frac{\lp^{-2}}{G} B_i'$ for some dimensionless coefficients $B_i'$. Summarizing the results in the previous paragraphs, we obtain for $H$, and hence for the operator $\hat{H}$,
\begin{align}
\label{eq:Hsummary}
H &= \frac{\lp^{-2}}{G} \sum_{i} \sum_{\sigma \in S_3}  \left(B_i' v^{n+1} e^{i \frac{\lp}{v} \sum_k (\sigma A_i')^k p_k c_k } + \cc \right) , \\
\label{eq:Hhatfinal}
\hat{H} &=  \frac{\lp^{-2}}{G} \sum_{i} \sum_{\sigma \in S_3} \left(
\reallywidehat{B_i' v^{n+1} e^{i \frac{\lp}{v} \sum_k (\sigma A_i')^k p_k c_k}} + \hc \right) + \BO(\lp^3) .
\end{align}

\subsubsection{Correct classical limit}
\label{subsubsec:corrclasslim}

Before we take the classical limit, we will simplify the expression for $H$ further. It is clear that in \eqref{eq:Hsummary} the action of the permutations $\sigma$ and complex conjugation generates terms differing only in coefficients $A_i^{'k}, B'_i$. Therefore, while the form (\ref{eq:Hsummary}) has the advantage of corresponding to the manifestly 
hermitian and reflection-invariant quantum operator, we can rewrite it in a simpler form as 
\beq
\label{eq:simp}
H = \frac{\lp^{-2}}{G} \sum^{N'}_{i=1}  B_i v^{n+1} e^{i \frac{\lp}{v} \sum_k A_i^k p_k c_k}
\eeq
for some $N'$ and suitably extended coefficients $A^k_i, B_i$. Now \eqref{eq:uniqf} yields that
\begin{equation}
\label{eq:uniqA}
\vec{A}_i = \vec{A}_j \text{ implies }i=j
\end{equation}
where $\vec{A}_i =(A_i^1, A_i^2, A_i^3)$.
In \eqref{eq:simp} the $B_i$ are equal for the terms related by permutations $\sigma$ and are complex conjugate of each other for the terms related by complex conjugation, while the set of coefficients $A^k_i$ is a disjoint union of subsets, where the elements $\vec{A}_i$  of each subset are related by the appropriate action of permutation and complex conjugation (negation) as in \eqref{eq:Hsummary}. 
Note that for pairs of terms related by \textit{both} a permutation and a complex conjugation, 
this implies in particular that the corresponding $B_i$'s are \textit{real}.
Furthermore, the form (\ref{eq:Hsummary}) for $H$ implies that the analogous expression for $\hat{H}$ holds
via the quantum-classical correspondence we have established:
\begin{equation}
\label{eq:hatHsimp}
\hat{H} = \frac{\lp^{-2}}{G}  \sum^{N'}_{i=1} B_i \reallywidehat{v^{n+1} e^{i\frac{\lp}{v} \sum_k A_i^k p_k c_k}} + \BO(\lp^3).
\end{equation}

We next expand the exponentials in powers of $\lp$:
\begin{displaymath}
H = \frac{\lp^{-2}}{G} \sum^{N'}_{i=1}  B_i v^{n+1} \left(1 + i\frac{\lp}{v} \sum_k A_{ik} p_k c_k - \frac{\lp^2}{2v^2} \sum_{k,l} A_{ik} A_{il} p_k p_l c_k c_l + \BO(\lp^3) \right),
\end{displaymath}
where we have collected coefficients $A^k_i$ into the matrix $A_{ik}:=A^k_i$.

Now we impose the condition that $H$ (\ref{eq:simp}) match the constraint $C_H$ \eqref{eq:hamc} in the classical limit. Let us repeat the classical constraint here to remind the reader of its form:
\begin{displaymath}
C_H  = -\frac{1}{8\pi G \gamma^2} v^{n-1} (p_1 p_2 c_1 c_2 + p_1 p_3 c_1 c_3 + p_2 p_3 c_2 c_3)
= - \frac{v^{n-1}}{2G} \sum_{ij} M^{ij} p_i p_j c^i c^j,
\end{displaymath}
where
\beq
M := \lambda \begin{pmatrix}
0 & 1 & 1 \\
1 & 0 & 1 \\
1 & 1 & 0
\end{pmatrix}
\eeq
and we set $\lambda=\frac{1}{8\pi \gamma^2}$.  The condition for the correct classical limit of $H$ then takes the form
\begin{align}
\nonumber
\lim_{\lp \rightarrow 0}
\frac{\lp^{-2}}{G} \sum^{N'}_{i=1}  B_i v^{n+1} \left(1 + i\frac{\lp}{v} \sum_k A_{ik} p_k c_k - \frac{\lp^2}{2v^2} \sum_{k,l} A_{ik} A_{il} p_k p_l c_k c_l + \BO(\lp^3) \right) \\
= - \frac{v^{n-1}}{2G} \sum_{i,j} M^{ij} p^i p^j c_i c_j .
\label{eq:classlim}
\end{align}
Note in particular that this condition implies that the terms with the negative powers of $\lp$ on the left-hand side must cancel. We obtain the conditions
\begin{align}
\label{eq:lpneg2cond}
&\sum_i \Re B_i = 0  \\
\label{eq:lpneg1cond}
&\sum_i A_{ij} \Im B_i = 0 \\
\label{eq:lp0cond}
&\sum_i A_{ij} \left(\Re B_i\right) A_{ik} = M_{jk}.
\end{align}
The matrix $A$ has three columns and $N'$ rows, corresponding to the coefficients $A^k_i$. Each row corresponds to a term in the expression \eqref{eq:simp} for $H$ . As mentioned above, the rows partition into sets related by permutations $\sigma$ and negation, and so are generated by some smaller, basic number of rows, one from each set.   

\subsubsection{Minimality and the simplest possibilities}

Now we introduce a key assumption in our analysis, which is meant to make precise the principle of `simplicity' (or Occam's razor) in the present case:

\textbf{Assumption 2 (minimality):} The number of terms $N'$ in $\hat{H}$ \eqref{eq:hatHsimp} is the smallest such that all of the other conditions on $\hat{H}$ stipulated
can be satisfied.

As we show below, the AW quantization of the Hamiltonian constraint corresponds to a matrix $A$ having 12 rows, generated by 
2 basic rows, and it satisfies 
all of the criteria which we impose on $\hat{H}$.  By Assumption 2, we therefore need only consider the case of $A$
having 12 rows or fewer. We will now list all of the possibilities for $A_{ij}$ and $B_i$, with 12 or fewer rows, satisfying all of the conditions so far (other than minimality). 

1. Matrix $A$ has eight rows generated by $\begin{pmatrix}
a_1 & a_1 & b_1 \\
a_2 & a_2 & a_2
\end{pmatrix}$ 
and $\vec{B} = (\b_1, \b_1, \b_1, \bar{\b}_1, \bar{\b}_1, \bar{\b}_1, \b_2, \bar{\b}_2)^T$.

Solutions are parametrized by three real numbers $a_1, b_1, c$ such that $a_1 \neq b_1$. The conditions imply
\begin{align*}
a_2 = \sqrt{\frac{2a_1^2+b_1^2}{3}} \quad\quad &\Re \b_1 = \frac{-\lambda}{2(a_1-b_1)^2} \\
\Re \b_2 = \frac{3\lambda}{2(a_1-b_1)^2} \quad\quad &\Im \b_1 = c \quad\quad \Im \b_2 = -\frac{2a_1+b_1}{a_2}c
\end{align*}

2. Matrix $A$ has eight rows generated by $\begin{pmatrix}
a_1 & -a_1 & 0 \\
a_2 & a_2 & a_2
\end{pmatrix}$
and $\vec{B} = (\b_1, \b_1, \b_1, \b_1, \b_1, \b_1, \b_2, \bar{\b}_2)^T$.

Solutions are parametrized by the real number $a_1$ such that $a_1 \neq 0$. The conditions imply
\begin{align*}
a_2 = \sqrt{\frac{2a_1}{3}} \quad\quad &\Re \b_1 = \frac{-\lambda}{6a_1^2} \\
\Re \b_2 = \frac{\lambda}{2a_1^2} \quad\quad &\Im \b_1 = 0 \quad\quad \Im \b_2 = 0
\end{align*}

3. Matrix $A$ has ten rows generated by $\begin{pmatrix}
a_1 & a_1 & b_1 \\
a_2 & a_2 & a_2 \\
a_3 & a_3 & a_3
\end{pmatrix}$ 
and $\vec{B} = (\b_1, \b_1, \b_1, \bar{\b}_1, \bar{\b}_1, \bar{\b}_1, \b_2, \bar{\b}_2, \b_3, \bar{\b}_3)^T$.

Solutions are parametrized by six real numbers $a_1, b_1, a_2, a_3, c_1, c_2$ such that $a_1 \neq b_1, a_2 \neq \pm a_3$. The conditions imply
\begin{align*}
&\Re \b_1 = \frac{-\lambda}{2(a_1-b_1)^2} \quad\quad \Re \b_2 = \frac{2a_1^2+b_1^2-3a_3^2}{a_3^2-a_2^2}\Re \b_1 \\
&\Re \b_3 = -3\Re \b_1 - \Re \b_2 \quad\quad \Im \b_1 = c_1 \\ 
&\Im \b_2 = c_2 \quad\quad \Im \b_3 = -\frac{(2a_1 + b_1)c_1 + a_2 c_2}{a_3} \quad \mathrm{if}\,\, a_3 \neq 0 \\
&\Im \b_3 = c_2 \quad\quad \Im \b_2 = -\frac{2a_1+b_1}{a_2}c_1  \quad \mathrm{if}\,\, a_3=0 
\end{align*}

4. Matrix $A$ has ten rows generated by $\begin{pmatrix}
a_1 & -a_1 & 0 \\
a_2 & a_2 & a_2 \\
a_3 & a_3 & a_3
\end{pmatrix}$
and $\vec{B} = (\b_1, \b_1, \b_1, \b_1, \b_1, \b_1, \b_2, \bar{\b}_2, \b_3, \bar{\b}_3)^T$.

Solutions are parametrized by four real numbers $a_1, a_2, a_3, c$ such that $a_1 \neq 0$, $a_2 \neq \pm a_3$. The conditions imply
\begin{align*}
&\Re \b_1 = \frac{-\lambda}{6a_1^2} \quad\quad \Re \b_2 = \frac{2a_1^2-3a_3^2}{a_3^2-a_2^2}\Re \b_1 \\ 
&\Re \b_3 = -3\Re \b_1 - \Re \b_2 \quad\quad \Im \b_1 = 0 \\ 
&\Im \b_2 = c \quad\quad \Im \b_3 = -\frac{a_2}{a_3}c \quad \mathrm{if}\,\, a_3 \neq 0 \\
&\Im \b_3 = c \quad\quad \Im \b_2 = 0  \quad \mathrm{if}\,\, a_3=0 
\end{align*}

5. \begin{minipage}[t]{5in} Matrix $A$ has twelve rows generated by $\begin{pmatrix}
a_1 & a_1 & b_1 \\
a_2 & a_2 & a_2 \\
a_3 & a_3 & a_3 \\
a_4 & a_4 & a_4
\end{pmatrix}$ \\
and $\vec{B} = (\b_1, \b_1, \b_1, \bar{\b}_1, \bar{\b}_1, \bar{\b}_1, \b_2, \bar{\b}_2, \b_3, \bar{\b}_3, \b_4, \bar{\b}_4)^T$.
\end{minipage}

\vspace{1em}

Solutions are parametrized by nine real numbers $a_1, b_1, a_2, a_3, a_4, c_1, c_2, c_3, d$ such that $a_1 \neq b_1$, 
$a_2 \neq \pm a_3$, $a_2 \neq \pm a_4$, $a_3 \neq \pm a_4$. The conditions imply
\begin{align*}
&\Re \b_1 = \frac{-\lambda}{2(a_1-b_1)^2} \quad\quad \Re \b_2 = d \quad\quad  \Re \b_3 = \frac{(2a_1^2+b_1^2-3a_3^2)\Re \b_1 + (a_2^2-a_4^2) \Re \b_2}{a_4^2-a_3^2} \\
&\Re \b_4 = -3\Re \b_1 - \Re \b_2 - \Re \b_3 \quad\quad \Im \b_1 = c_1  \quad\quad \Im \b_2 = c_2 \\ 
&\Im \b_3 = c_3 \quad\quad \Im \b_4 = -\frac{(2a_1 + b_1)c_1 + a_2 c_2 + a_3 c_3}{a_4} \quad \mathrm{if}\,\, a_4 \neq 0 \\
&\Im \b_4 = c_3 \quad\quad \Im \b_3 = -\frac{(2a_1+b_1)c_1 + a_2 c_2}{a_3}  \quad \mathrm{if}\,\, a_4=0 
\end{align*}

6. \begin{minipage}[t]{5in} Matrix $A$ has twelve rows generated by $\begin{pmatrix}
a_1 & -a_1 & 0 \\
a_2 & a_2 & a_2 \\
a_3 & a_3 & a_3 \\
a_4 & a_4 & a_4
\end{pmatrix}$ \\
and $\vec{B} = (\b_1, \b_1, \b_1, \b_1, \b_1, \b_1, \b_2, \bar{\b}_2, \b_3, \bar{\b}_3, \b_4, \bar{\b}_4)^T$.
\end{minipage}

\vspace{1em}

Solutions are parametrized by seven real numbers $a_1, a_2, a_3, a_4, c_2, c_3, d$ such that $a_1 \neq 0$, 
$a_2 \neq \pm a_3$, $a_2 \neq \pm a_4$, $a_3 \neq \pm a_4$. The conditions imply
\begin{align*}
&\Re \b_1 = \frac{-\lambda}{6a_1^2} \quad\quad \Re \b_2 = d \quad\quad \Re \b_3 = \frac{(2a_1^2-3a_4^2) \Re \b_1 + (a_2^2 - a_4^2) \Re \b_2}{a_4^2-a_3^2} \\ 
&\Re \b_4 = -3\Re \b_1 - \Re \b_2 - \Re \b_3 \quad\quad \Im \b_1 = 0 \quad\quad \Im \b_2 = c_2 \\ 
&\Im \b_3 = c_3 \quad\quad \Im \b_4=-\frac{a_2 c_2 + a_3 c_3}{a_4} \quad \mathrm{if}\,\, a_4 \neq 0 \\
&\Im \b_4 = c_3 \quad\quad \Im \b_3 = -\frac{a_2}{a_3}c_2  \quad \mathrm{if}\,\, a_4=0 
\end{align*}

7. \begin{minipage}[t]{5in}Matrix $A$ has twelve rows generated by  $\begin{pmatrix}
a_1 & -a_1 & 0 \\
a_2 & a_2 & b_2 \\
\end{pmatrix}$\\
and $\vec{B} = (\b_1, \b_1, \b_1, \b_1, \b_1, \b_1, \b_2, \b_2, \b_2, \bar{\b}_2, \bar{\b}_2, \bar{\b}_2)^T$.
\end{minipage}

\vspace{1em}

Solutions are parametrized by two real numbers $a_2, b_2$ such that $b_2 \neq -2a_2$. The conditions imply
\begin{align*}
a_1 = \sqrt{\frac{2a_2^2+b_2^2}{2}} \quad\quad &\Re \b_1 = \frac{-\lambda}{(2a_2+b_2)^2} \\
\Re \b_2 = -\Re \b_1 = \frac{\lambda}{(2a_2+b_2)^2} \quad\quad &\Im \b_1 = 0 \quad\quad \Im \b_2 = 0
\end{align*}

8. \begin{minipage}[t]{5in} Matrix $A$ has twelve rows generated by  $\begin{pmatrix}
a_1 & a_1 & b_1 \\
a_2 & a_2 & b_2 \\
\end{pmatrix}$\\
and $\vec{B} = (\b_1, \b_1, \b_1, \bar{\b}_1, \bar{\b}_1, \bar{\b}_1, \b_2, \b_2, \b_2, \bar{\b}_2, \bar{\b}_2, \bar{\b}_2)^T$.
\end{minipage}

\vspace{1em}

Then the solutions fall into two classes:

a) solutions parametrized by four real numbers $a_1, a_2, b_1, c$ such that $b_2 \neq -2a_2$, 
$2a_1^2+b_1^2 \geq 2a_2^2$, $a_1^2+2a_1 b_1 \neq a_2^2 + 2a_2 b_2$. The conditions imply
\begin{align*}
b_2 = \pm\sqrt{2(a_1^2-a_2^2)+b_1^2} \quad\quad &\Re \b_1 = \frac{\lambda}{2\left(a_1^2-a_2^2+2a_1b_1 \mp 2a_2\sqrt{2(a_1^2-a_2^2)+b_1^2}\right)} \\
\Re \b_2 = -\Re \b_1  \quad\quad &\Im \b_1 = c \quad\quad \Im \b_2 = -\frac{2a_1+b_1}{2a_2+b_2}c
\end{align*}

b) solutions parametrized by three real numbers $a_1, b_1, c$ such that $b_1 \neq 2a_1$. The conditions imply
\begin{align*}
a_2 = \sqrt{\frac{2a_1^2+b_1^2}{6}} \quad\quad &b_2 = -2a_2 \quad\quad \\  \Re \b_1 = \frac{\lambda}{2(2a_1^2+b_1^2)} \quad\quad &\Re \b_2 = -\Re \b_1  \\  \Im \b_2 = c \quad\quad &\Im \b_1 = 0
\end{align*}

\subsection{Planar loops}

We will now impose a physical assumption that arises from the fact that $\hat{H}$
should be the quantization of a classical expression involving local fields.
Specifically, we will require that the curvature is obtained by taking holonomies of the connection around planar loops. This translates to the condition that every row of the matrix $A$ contain a zero. By considering the solutions above, only the family of solutions (7) is able to satisfy this condition. 
Solution (7) is represented by a matrix $A$ that has twelve rows generated by  
$\begin{pmatrix}
a_1 & -a_1 & 0 \\
a_2 & a_2 & b_2 \\
\end{pmatrix}$.
The planar loops condition imposes $b_2 = 0$. The resulting matrix $A$
is then exactly the matrix required to obtain the AW Hamiltonian.

Indeed, by the planar loops condition $b_2 = 0$. It follows that $a_2$ is the only free parameter. 
This is consistent with what usually happens when defining dynamics in LQC:
there, too, the $\bar{\mu}$ parameter is not uniquely determined and requires input from the full theory. Thus, this remaining freedom was expected. For the AW case the parameter $a_2$ equals $\sqrt{\Delta}$ with $\Delta \lp^2$ being the minimum eigenvalue of the area operator. Then, by using solution (7) from the last subsection, we get
\begin{displaymath}
a_1 = \sqrt{\Delta} \quad\mathrm{and}\quad
-\b_1 = \b_2 = \frac{\lambda}{4\Delta} = \frac{1}{32\pi \gamma^2 \Delta} .
\end{displaymath}

The AW Hamiltonian is given by \cite{aw2009a}
\begin{align*}
H_{AW} = -\frac{1}{8\pi G \gamma^2 \Delta \lp^2} \Big(&p_1 p_2 |p_3| \sin(\bar{\mu}_1 c_1) \sin(\bar{\mu}_2 c_2) + \\ &+
|p_1| p_2 p_3 \sin(\bar{\mu}_3 c_3) \sin(\bar{\mu}_2 c_2) + p_1 |p_2| p_3 \sin(\bar{\mu}_1 c_1) \sin(\bar{\mu}_3 c_3)\Big)
+ \BO(\lp^3),
\end{align*}
where $\bar{\mu}_1=\sqrt{\frac{|p_1|\Delta\lp^2}{|p_2p_3|}}=\frac{\sqrt{\Delta}\lp}{v}|p_1|$ and other $\bar{\mu}_i$ are defined by cyclic permutations. By writing the sines as exponentials and using the identity \eqref{eq:parident} together with the BCH formula\footnote{Though
the action of the shift operators $\widehat{e^{i\vec{f}(\vec{p})\cdot \vec{c}}}$ \eqref{starhom} on the Bohr Hilbert space is not
the operator exponential of the quantization of $\vec{f}(\vec{p})\cdot \vec{c}$ (which doesn't exist on the Bohr Hilbert space),
its action on the Schr\"odinger Hilbert space $L^2(\mathbb{R}^3) \ni \psi(\vec{p})$ \textit{is} 
such an operator exponential \cite{aps2006}, so that the usual BCH formula applies.
},
we obtain
\begin{align*}
H_{AW} = & \frac{v^2}{32\pi G \gamma^2 \Delta \lp^2} \Big(e^{i\left(\frac{\sqrt{\Delta}\lp}{v} (p_1 c_1 + p_2 c_2)\right)} - e^{i\left(\frac{\sqrt{\Delta}\lp}{v} (p_1 c_1 - p_2 c_2)\right)} + e^{i\left(\frac{\sqrt{\Delta}\lp}{v} (p_2 c_2 + p_3 c_3)\right)} - \\ &- e^{i\left(\frac{\sqrt{\Delta}\lp}{v} (p_2 c_2 - p_3 c_3) \right)} + e^{i\left(\frac{\sqrt{\Delta}\lp}{v} (p_1 c_1 + p_3 c_3)\right)} - e^{i\left(\frac{\sqrt{\Delta}\lp}{v} (p_1 c_1 - p_3 c_3) \right)} + \hc\Big) + \BO(\lp^3),
\end{align*}
which matches the solution we found above for $n=1$.

\section{Projection to isotropic LQC}
\label{sec:proj}

In \cite{aw2009a} the authors define a projector $\hat{\mathbb{P}}$ from the states of the Bianchi I model to the states of the isotropic model. The projector $\hat{\mathbb{P}}$ acts on the states $\Psi(p_1, p_2, v)$ in the Bianchi I model and projects them down to the states $\psi(v)$ in the Friedmann model \cite{aw2009a}:
\begin{displaymath}
(\hat{\mathbb{P}} \Psi) (v) := \sum_{p_1, p_2} \Psi(p_1, p_2, v) \equiv \psi(v),
\end{displaymath}
which is equivalent to
\begin{displaymath}
\hat{\mathbb{P}} |p_1,p_2,v \rangle = |v \rangle .
\end{displaymath}
Applying this projector to the Bianchi I Hamiltonian, the authors obtain the Hamiltonian for the Friedmann model that exactly reproduces the one introduced in \cite{aps2006}. Because we derived in the previous section the AW Hamiltonian, our result will project to the APS Hamiltonian in exactly the same manner.

However, we can relax one of the assumptions that led us to the unique form of the Bianchi I Hamiltonian and project to the isotropic cosmology, thus providing an alternative derivation for the APS Hamiltonian. We will find that \textit{the planar loops assumption is not needed to obtain the Hamiltonian for the isotropic model.} 

To see this, we first note that,
according to assumption 1, each $\F_i$ is generated as a flow and therefore is a solution to the initial value problem
\begin{displaymath}
\frac{d}{dt} \F^t_i(\p) = 8\pi \gamma G\hbar \f\left(\F^t_i(\p)\right) \quad\quad \F^0_i(\p) = \p.
\end{displaymath}
Using the results of the previous section, we have
\begin{displaymath}
f^k_i (\p) = \lp A_i^k \frac{p^k}{v}.
\end{displaymath}
Therefore, the initial value problem can be rewritten as 
\begin{displaymath}
\frac{d}{dt} (F^t_i(\p))^k = 8\pi \gamma \lp^3 A_i^k \frac{(F^t_i(\p))^k}{\sqrt{\prod_k |(F^t_i(\p))^k|}} \quad\quad (F^0_i(\p))^k = p^k.
\end{displaymath}

One can check that $\tilde{F}^t_i(\p)^k := p^k(1+8\pi \gamma t A_i^k \lp^3/v)$
solves these conditions up to terms of order subleading by at least $\BO(\lp^3)$. Thus
\begin{displaymath}
(F^t_i(\p))^k = p^k \left(1+ 8\pi \gamma t A_i^k \frac{\lp^3}{v} + \BO(\lp^6)\right).
\end{displaymath}
Changing variables, we obtain that each shift operator acts as 
\begin{equation}
\label{eq:BIshift}
\widehat{e^{i \f_i(\p) \cdot \c}} |p_1, p_2, v\rangle= |(F_i^1(\p))^1, (F_i^1(\p))^2, v'\rangle,
\end{equation}
where $v'$ is given by
\begin{equation}
\label{eq:vprime}
v' = \sqrt{\prod_k \left\lvert (F^t_i(\p))^k \right\rvert} 
= v\left(1+ 4\pi \gamma \frac{\lp^3}{v}\sum_k A^k_i + \BO(\lp^6)\right) .
\end{equation}
Note that, in addition to its present purpose, this equation allows us to calculate the $v_0$ defined in section \ref{subsubsec:parity}
to be $v_0 = 4\pi \gamma \sup_i |\sum_k A^k_i| \lp^3$.
The map $v \mapsto v'$ is equivalent to the map
\begin{displaymath}
p \mapsto p' = p\left(1+ \frac{8\pi \gamma \lp^3}{3 v}\sum_k A^k_i + \BO(\lp^6)\right)
\end{displaymath}
which is generated by the vector field
\begin{displaymath}
\frac{8\pi \gamma G \hbar}{3}\left(\frac{\lp p}{v}\sum_k A^k_i + \BO(\lp^4)\right)\frac{d}{dp},
\end{displaymath}
so that the operator mapping $|v\rangle$ to $|v'\rangle$ in the isotropic theory \cite{aps2006, abl2003}) is
\begin{displaymath}
\reallywidehat{e^{i \lp (\sum_k A^k_i)\frac{pc}{v} + \BO(\lp^4)}} = 
\reallywidehat{e^{i \lp (\sum_k A^k_i)\frac{pc}{v}}} + \BO(\lp^4) .
\end{displaymath}

 Equations (\ref{eq:BIshift}) and (\ref{eq:vprime}) thus imply
\begin{displaymath}
\hat{\mathbb{P}} \circ \widehat{e^{i \f_i(\p) \cdot \c}} = 
\reallywidehat{e^{i \lp (\sum_k A^k_i)\frac{pc}{v}}} \circ \hat{\mathbb{P}} + \BO(\lp^4).
\end{displaymath}
Furthermore, for any function $g(v)$, $\hat{\mathbb{P}} \circ g(v) = g(v) \circ \hat{\mathbb{P}}$. Therefore, if we start from the Hamiltonian (\ref{eq:hatHsimp}), the unique Hamiltonian $\hat{H}_{FLRW}$ in the isotropic model satisfying $\hat{H}_{FLRW} \circ \hat{\mathbb{P}} =\hat{\mathbb{P}} \circ \hat{H}$ is given by
\begin{displaymath}
\hat{H}_{\mathrm{FLRW}} = \frac{\lp^{-2}}{G} \sum^{N'}_{i=1}  B_i \reallywidehat{v^{n+1}  e^{i \frac{\lp}{v} (\sum_k A_i^k) p c}} + \BO(\lp^3),
\end{displaymath}
where we choose in the isotropic theory the same ordering convention as for the Bianchi I model in subsection \ref{subsec:fixclassan}.
This implies that the analysis of the subsection \ref{subsec:classlim} carries through and the final possibilities (1)-(8) for  the matrix $A$ are then transformed into column vectors by summing each row $A_i = \sum_k A_i^k$. Another assumption we make is the assumption of the minimum number of terms. By considering each of the possibilities (1)-(8) outlined above, it is clear that the minimum number of terms is three, corresponding to the column vector $A_i = (0, a, -a)^T$. 

Now, the isotropic Hamiltonian given by \cite{aps2006} is
\bq
\hat{H}_{APS} = \sin(\bar{\mu}c) \left[ \frac{24i \sgn(p)}{8\pi \gamma^3 \bar{\mu}^3 \lp^2} \left(\sin\left(\frac{\bar{\mu}c}{2}\right)\hat{V}\cos\left(\frac{\bar{\mu}c}{2}\right) - \cos\left(\frac{\bar{\mu}c}{2}\right) \hat{V}\sin\left(\frac{\bar{\mu}c}{2}\right)\right)\right] \sin(\bar{\mu}c),
\eq
where $\bar{\mu} = \sqrt{\frac{\Delta \lp^2}{|p|}}$, and $\hat{V}$ is the volume operator.
To leading and subleading order in $\lp$ it can be written as 
\bq
\hat{H}_{APS} = \frac{3}{2\Delta \BI^2 \lp^{2}}  v \left(-2 + e^{i\left(\frac{\lp}{v}2\sqrt{\Delta}pc\right)} + e^{-i\left(\frac{\lp}{v}2\sqrt{\Delta}pc\right)}\right) + \BO(\lp^3),
\eq
which matches the solution we found above for $a=2\sqrt{\Delta}$ and $n=0$.\footnote{In \cite{aps2006} the authors define the classical constraint $C_H = -\frac{6}{\BI^2} c^2 \sqrt{p}$. To match this constraint the constant $\lambda$ introduced in \ref{subsubsec:corrclasslim} must be equal to $2G\BI^{-2}$.}

\section{Discussion}
\label{sec:disc}

In this paper we derived the Hamiltonian constraint operator for the Bianchi I cosmology. The result matches the form previously proposed by Ashtekar and Wilson-Ewing in \cite{aw2009a}. We have also obtained the quantum Hamiltonian for the FLRW model which agrees with the one given in \cite{aps2006}. Thus, our work provides an independent derivation for the quantum Hamiltonian obtained in those papers, increasing confidence in LQC models.

The ingredients for our derivation are basic physical principles, such as diffeomorphism invariance, and certain simplifying assumptions. We start by writing down the  general form for the Hamiltonian as an operator that preserves the Hilbert space of states dictated by the use of the holonomy-flux algebra \cite{ac2012,eh2016, eht2016}. This operator is required to be hermitian and have as its classical analogue a function on the classical phase space. The latter condition enables us to study the classical limit of the quantum Hamiltonian in a state-independent way (without considering its expectation value on semi-classical states). 
We then proceed to constrain its form by imposing covariance under the residual diffeomorphism symmetries 
--- more specifically invariance under the canonical ones and covariance under the non-canonical ones.
Matching the classical limit of the quantum Hamiltonian to the classical Hamiltonian constraint, we arrive at a discrete set of finite-dimensional families of possibilities.
Finally, we use an input from the full theory, namely the quantization of curvature by holonomies around planar loops, together with a minimality principle.
This yields a unique form of the Hamiltonian constraint parametrized by a single parameter which exactly corresponds to the Hamiltonian in \cite{aw2009a},
with the single parameter corresponding to the area gap (which also in \cite{aw2009a} must be `parachuted in' from the full theory).
Furthermore, by projecting down to the isotropic model we obtain uniquely the `improved dynamics' form of the Hamiltonian proposed in \cite{aps2006} \textit{without recourse to the planar loops assumption}.

A crucial step in this derivation of the Hamiltonian constraint is the requirement of covariance under volume-changing dilations.
Because volume-changing dilations do not preserve the symplectic structure of the classical phase space, they are not well-defined as unitary operators on states in the quantum theory.  Nevertheless, an action of such dilations can be defined \textit{on operators}, unique up to 
ordering ambiguity. This ordering ambiguity in the definition of volume-changing dilations leads to a corresponding ambiguity in the Hamiltonian constraint --- the only ambiguity which cannot be fixed with the principles considered. Fortunately, this ambiguity turns out to affect neither the leading nor subleading order terms, in $\lp$, of the resulting Hamiltonian constraint. 
Only the leading and subleading order terms are relevant for the effective equations \cite{aps2006}, which, in the isotropic case, have been found to be an accurate reflection of the exact quantum theory \cite{djms2017, dgms2014, dgs2014} to a sufficient extent that it is the effective equations which are used in the calculation of the distribution of primordial perturbations predicted by loop quantum cosmology \cite{aan2013, aan2013a, agullo2015, abs2017}. In fact, even if one were to include the sub-subleading order terms in the effective equations, it is likely that such terms yield a sub-subleading correction to the predicted power spectrum \cite{aan2013}. In this latter work, the evolution equations for the Fock modes on the quantum geometry background were shown to depend on only two ``moments'' 
$\langle \hat{H}_o^{-1}\rangle$ and $\langle \hat{H}_0^{-1/2} \hat{a}^4 \hat{H}_o^{-1/2}\rangle$, where $\hat{H}_o = \hbar \sqrt{\hat{H}}$. It appears that the sub-subleading, in $\lp$, terms in $\hat{H}$ would give only sub-subleading corrections to these moments, and therefore to the evolution of the modes and prediction of the power spectrum. 

To summarize,  the present work shows that, beyond physical principles, the only choices required in the derivation of the
Hamiltonian of isotropic LQC are (1.)  the use of the holonomy-flux algebra and (2.) minimality; in the Bianchi I case, the only further assumption required is that of planar loops. Therefore, for the isotropic case, the present work, with \cite{ac2012, eh2016, eht2016}, shows that \textit{predictions based on the standard APS Hamiltonian are in fact predictions coming only from the use of the holonomy-flux algebra, the core assumption in LQG, together with minimality.} In particular, assuming minimality, this paper supports an even higher confidence in the power spectrum
\cite{aan2013, aan2013a} and bispectrum \cite{abs2017} as predictions of LQG itself.

It is remarkable that in the isotropic case all (physically relevant) ambiguity in the Hamiltonian constraint can be eliminated with a single assumption --- the minimality principle. Nevertheless, it is interesting to relax this assumption and investigate whether the remaining possibilities lead to qualitatively different predictions. This has already begun to be analyzed \cite{lsw2018, adl2018} for one of the non-minimal Hamiltonians selected
in this paper, which has also been proposed in the literature before \cite{ydm2009, ld2017}, motivated by quantization techniques more closely imitating those originally proposed by Thiemann for the Hamiltonian constraint in full loop quantum gravity. 
In these works, the effective equations for the quantum background geometry, and solutions thereof, are determined and analyzed.
What remains is to determine whether this alternative Hamiltonian also leads to a different distribution of primordial perturbations.  Such analysis would perhaps also be interesting for other non-minimal Hamiltonians selected in the present paper.

\section*{Acknowledgements}

The authors thank Abhay Ashtekar, Christopher Beetle, and Muxin Han for discussions, and thank Martin Bojowald for emphasizing the need to extend uniqueness theorems in LQC based on covariance from kinematics to dynamics.
This work was supported in part by NSF grant PHY-1505490.

\appendix
\section*{Selecting Hamiltonian operator without using non-canonical dilations}

This appendix provides an alternative argument for \eqref{eq:Hsummary} which does not require use of non-canonical dilations. Instead it uses a minimal input from quantization (a small part of what was originally used by Ashtekar and Wilson-Ewing in \cite{aw2009a}).

Starting from \eqref{eq:Hdisc} and similar to \ref{subsec:classlim}, we require that the Hamiltonian reduces to the classical constraint in the classical limit. Introducing the dependence of $\hat{H}$ on the classicality parameter $\lp$,
\beq
\label{eq:ham_app}
\hat{H} = \sum_{i} \sum_{\sigma \in S_3} \left( e^{i \sum_k (\sigma \tf_i)^k(v, \lp) p^k c_k}g_i(v,\lp) + \hc \right).
\eeq
Replacing $\hat{H}$ by its classical analogue $H$ we get
\begin{equation}
\label{eq:Hclassan_app}
H = \sum_{i} \sum_{\sigma \in S_3} \left( e^{i \sum_k (\sigma\tf_i)^k(v,\lp) p^k c_k}g_i(v,\lp) + \cc \right).
\end{equation}

We will now use crucially an input from quantization. Namely, we assume that curvature entering the classical Hamiltonian constraint is quantized using holonomies around loops. Furthermore, these loops are chosen to have either area \cite{rs1995} or length \cite{thiemann1996a} described 
by a minimum quantum number. The length of any part of the loop is then proportional to the Planck length $\lp$, and therefore $\tf^k_i(v,\lp) = \lp \tf^k_i(v)$. Given that $c_i$ is dimensionless and $p_i$ has the dimension of an area, it is clear that $\tf^k_i(v,\lp)$ has the dimension of an inverse area.  Requiring (as in the main text) that the \textit{only length scale in the theory} be the \textit{Planck length} $\lp$ leads us to conclude that\footnote{Even without quantization assumptions, one can show by matching $H$ to the classical expression that $\tf^k_i(v,\lp) = \BO(\lp)$.}
\beq
\label{eq:ftilde}
\tf^k_i(v,\lp) = \tilde{A}^k_i \frac{\lp}{v}
\eeq
with $\tilde{A}^k_i$ a constant. 

Turning now to the functions $g_i(v,\lp)$, we again note that, since $H$ equals the Hamiltonian constraint in the classical limit, the dimension of $g_i(v,\lp)$ should match the dimension of the classical constraint \eqref{eq:hamc}. This yields $g_i(v,\lp) = \frac{\lp^{3n+1}}{G} \tilde{g}_i(v,\lp)$ for some dimensionless function $\tilde{g}(v,\lp)$, taken to be analytic. The  requirement that $\lp$ be the only length scale in the theory implies that the coefficients in the two-variable Laurent expansion of $\tilde{g}_i(v, \lp)$ have to be dimensionless and, therefore, we can write
\beq
\label{eq:gexp}
g_i(v,\lp) = \frac{\lp^{3n+1}}{G} \left(\sum^{\infty}_{j=j_0} \tilde{B}^j_i \frac{\lp^{3j}}{v^j}\right),
\eeq
where for future convenience we denote the lower bound of the Laurent series by $j_0$ (which for the moment is arbitrary including $-\infty$), and $\tilde{B}^j_i$ are complex coefficients.

Summarizing the results in the previous paragraphs, we obtain for the operator $\hat{H}$ and its classical analogue $H$,
\begin{align}
\label{eq:Hhatfinalapp}
\hat{H} &= \frac{\lp^{3n+1}}{G} \sum_{i} \sum_{\sigma \in S_3} 
e^{i \left(\frac{\lp}{v} \sum_k (\sigma\tilde{A}_i)^k p_k c_k\right)}
\left(\sum^{\infty}_{j=j_0} \tilde{B}^j_i \frac{\lp^{3j}}{v^j}\right) + \hc ,
\\
\label{eq:Hsummaryapp}
H &=  \frac{\lp^{3n+1}}{G}  \sum_{i} \sum_{\sigma \in S_3}
\left(\sum^{\infty}_{j=j_0} \tilde{B}^j_i \frac{\lp^{3j}}{v^j}\right) 
e^{i \left(\frac{\lp}{v} \sum_k (\sigma\tilde{A}_i)^k p_k c_k \right)} +\cc .
\end{align}
Here the action of the permutations $\sigma$ and complex conjugation generates terms differing only in coefficients $\tilde{A}_i^k, \tilde{B}^j_i$, so that we can rewrite $H$ in the simpler form
\beq
\label{eq:simpapp}
H = \frac{\lp^{3n+1}}{G} \sum^{N'}_{i=1} \left(\sum^{\infty}_{j=j_0} B^j_i \frac{\lp^{3j}}{v^j} \right) e^{i \left(\frac{\lp}{v} \sum_k A_i^k p_k c_k\right)} 
\eeq
for some $N'$ and suitably extended set of coefficients $A^k_i, B^j_i$.  \eqref{eq:uniqf} yields that
\begin{equation}
\label{eq:uniqAapp}
\vec{A}_i = \vec{A}_j \text{ implies }i=j,
\end{equation}
where $\vec{A}_i =(A_i^1, A_i^2, A_i^3)$.
In \eqref{eq:simpapp} the $B^j_i$ are equal for the terms related by permutations $\sigma$ and are complex conjugate of each other for the terms related by complex conjugation, while the set of coefficients $A^k_i$ is a disjoint union of subsets, where the elements $\vec{A}_i$  of each subset are related by the appropriate action of permutation and complex conjugation. 
The quantum operator $\hat{H}$ can be cast in analogous form:
\begin{equation}
\label{eq:hatHsimpapp}
\hat{H} =  \frac{\lp^{3n+1}}{G} \sum^{N'}_{i=1} \sum^{\infty}_{j=j_0} B^j_i \reallywidehat{\frac{\lp^{3j}}{v^j} e^{i \left(\frac{\lp}{v} \sum_k A_i^k p_k c_k\right)}}.
\end{equation}

We next expand the exponentials in powers of $\lp$:
\begin{displaymath}
H =  \frac{\lp^{3n+1}}{G} \sum^{N'}_{i=1}\left(\sum^{\infty}_{j=j_0} B^j_i \frac{\lp^{3j}}{v^j}\right) \left(1 + i\frac{\lp}{v} \sum_k A_{ik} p_k c_k - \frac{\lp^2}{2v^2} \sum_{k,l} A_{ik} A_{il} p_k p_l c_k c_l + \BO(\lp^3) \right),
\end{displaymath}
where we have collected coefficients $A^k_i$ into the matrix $A$.
Next, we return to the condition that $H$ match the Hamiltonian constraint in the classical limit. To ensure that the classical limit does not blow up, the terms with the negative powers of $\lp$ again must cancel. Since only the terms with the same power of $c$ (denoted by $m$) can cancel, and the power of $\lp$ is $3n+1+3j+m$, it follows that only terms with coefficients $B^j_i$ with the same $j$ can cancel. Therefore, it is the terms with coefficients $B^j_i, j<-n-1$ that produce cancellations. Thus, we can set $B^j_i=0$ for $j<-n-1$ without changing $H$, and therefore without changing $\hat{H}$ up to operator ordering ambiguities, that is, up to $\BO(\lp^3)$ (see section \ref{subsec:classan}). Furthermore, this will not affect the equations for $B^j_i$ with other $j$ because they are only coupled via $A_{ik}$. Thus, without loss of generality we can set $j_0=-n-1$:
\beq
H =  \frac{\lp^{3n+1}}{G} \sum^{N'}_{i=1} \left(\sum^{\infty}_{j=-n-1} B^j_i \frac{\lp^{3j}}{v^j}\right) \left(1 + i\frac{\lp}{v} \sum_k A_{ik} p_k c_k - \frac{\lp^2}{2v^2} \sum_{k,l} A_{ik} A_{il} p_k p_l c_k c_l + \BO(\lp^3) \right) .
\eeq

The classical Hamiltonian constraint is quadratic in $c$ and therefore corresponds to the terms in the expansion of $g$ \eqref{eq:gexp} with $j=-n-1$. The terms with coefficients $B^j_i, j>-n-1$ can not be constrained by imposing the classical limit as they are higher order in $\lp$. At this point, for this alternative argument, we use a stronger version of the minimality criterion in the main text (so that this criterion replaces and implies the minimality condition in \ref{subsec:classlim}). Specifically, we assume that the number of terms in \eqref{eq:hatHsimpapp}, in the sum over \textit{both} $i$ and $j$, is the smallest required to satisfy all of the other conditions on $\hat{H}$. This assumption implies that the coefficients $B^j_i$ with $j>-n-1$ are zero, leaving the coefficients $B^j_i$ with $j=-n-1$, which are precisely the coefficients $B_i$ in \eqref{eq:simp}. 
The expression \eqref{eq:hatHsimpapp} for $\hat{H}$ then reduces to expression \eqref{eq:hatHsimp} for $\hat{H}$, and the argument proceeds from there as in the main text.


%
\end{document}